\documentclass[10pt]{article}
\usepackage{makeidx}
\usepackage{multirow}
\usepackage{multicol}
\usepackage{graphicx}
\usepackage{epstopdf}
\usepackage{ulem}
\usepackage{hyperref}
\usepackage{amsmath}
\usepackage{amssymb}
\usepackage{setspace}
\usepackage{color}
\usepackage{float}
\author{Aishwarjya Gogoi}
\title{}
\usepackage[paperwidth=612pt,paperheight=792pt,top=72pt,right=72pt,bottom=72pt,left=72pt]{geometry}

\makeatletter
	{\par\setlength{\parindent}{#3}
	\setlength{\leftmargin}{#1}       \setlength{\rightmargin}{#2}%
	\advance\linewidth -\leftmargin       \advance\linewidth -\rightmargin%
	\advance\@totalleftmargin\leftmargin  \@setpar{{\@@par}}%
	\parshape 1\@totalleftmargin \linewidth\ignorespaces}{\par}%
\makeatother 

\begin{document}
\begin{center}
\textbf{{\LARGE A simple HLLE-type scheme for all Mach number flows}}
\end{center}

\begin{center}
A. Gogoi and  J.C. Mandal

*Department of Aerospace Engineering, Indian Institute of Technology, Bombay,Mumbai, 400076, India
\end{center}

\begin{doublespace}
\begin{center}
\uppercase{ABSTRACT}
\end{center}

A simple HLLE-type scheme  is proposed  for all Mach number flows. In the proposed scheme, no extra wave structure is added in the HLLE scheme to resolve the shear wave while the contact wave  is resolved by adding a wave structure similar to the HLLEM scheme. The resolution of the shear layers and the flow features at low Mach number are achieved by a velocity reconstruction method based on the face normal Mach number. Robustness against the numerical instabilities is achieved by scaling the velocity reconstruction function in the vicinity of shock with a multi-dimensional pressure sensor. The ability of the proposed scheme to resolve  low Mach flow features is demonstrated through asymptotic analysis while stability  of the proposed scheme for strong shock is demonstrated through  linear perturbation and matrix stability analyses. A set of numerical test cases are solved to show that the proposed scheme is free from numerical shock instability problems at high speeds and is capable of resolving the flow features at very low Mach numbers.


\section{Introduction }
The Godunov-type approximate Riemann solvers \cite{godu, toro-book} are popular  methods  for computing the  convective fluxes in the Euler and the Navier-Stokes equations. The approximate Riemann solvers are broadly classified into incomplete and complete solvers. The incomplete approximate Riemann  solvers do not incorporate all the linearly degenerate and non-linear waves that are present in the Riemann problem for evaluating the numerical flux. Whereas, the complete solvers include all the waves. The HLL \cite{hll} and the HLLE schemes \cite{einf1} are examples of incomplete approximate Riemann solvers while  examples of complete approximate Riemann solvers are the Roe scheme \cite{roe}, the Osher scheme \cite{osher}, the HLLEM scheme \cite{einf2} and the HLLC scheme  \cite{toro1, batt}. The HLL scheme \cite{hll}, proposed by Harten, Lax and van Leer,  satisfies entropy conditions and resolves isolated shock exactly. The HLL scheme with the wave speed estimates of Einfeldt \cite{einf1}, commonly known as the HLLE scheme,  possess entropy satisfying and  positivity preserving properties desirable for a numerical scheme.  However, the HLL scheme assumes an incomplete, two-wave structure of the Riemann problem comprising of only the non-linear acoustic waves and hence is incapable of resolving the contact and shear waves. The HLL scheme is also known to generate non-physical results in low Mach number flows. Nevertheless, due to the desirable entropy satisfying and positivity preserving properties, the HLL framework is highly preferred and several approaches  have been developed to introduce the contact and shear waves into the HLL framework.  The most popular schemes based on the HLL framework are HLLEM by Einfeldt et al. \cite{einf2} and HLLC by Toro et al. \cite{toro1}.

In the HLLEM scheme  proposed by Einfeldt et al. \cite{einf2}, anti-diffusion terms are added to the HLLE scheme corresponding to the linearly degenerated contact and shear waves that are not resolved in the HLL scheme. In this process, the HLLEM scheme achieves the accuracy  similar to the Roe scheme \cite{roe}. It is entropy satisfying and positively conservative like the HLLE scheme.  In the HLLC scheme  proposed by Toro et al. \cite{toro1}, the contact and the shear waves  are restored in the HLL scheme explicitly by enforcing the Rankine-Hugoniot jump conditions. The HLLC scheme  preserves the entropy satisfying and positively conservative properties of the HLLE scheme. The HLLC scheme with the wave-speed estimate  of Batten et al. \cite{batt} has emerged as perhaps the most preferred scheme due to its ability to resolve the linearly degenerate discontinuities. However,  for high speed flows, both the  HLLEM \cite{einf2} and  HLLC \cite{toro1, batt} schemes  are susceptible to numerical shock instabilities like odd-even decoupling, kinked Mach stem,  carbuncle phenomenon and low frequency post-shock oscillations. Even the exact Riemann solver of Godunov \cite{godu} and the other complete approximate Riemann solvers like the Roe  \cite{roe} and Osher  \cite{osher} schemes  also suffer from these numerical shock instability problems. On the other hand, the HLLE scheme \cite{hll, einf1} which do not resolve the linearly degenerate waves, is free from  numerical shock instabilities. The cause of numerical instabilities in the complete Riemann solvers have been investigated by several authors ever since the numerical instabilities  were first presented by Peery and Imlay \cite{peery}.  Quirk has carried out  linearized perturbation analysis \cite{quirk} and has observed that  the schemes in which the pressure perturbations feed the density perturbations is  afflicted by the odd-even decoupling problem. Quirk has proposed a hybrid Roe-HLLE scheme \cite{quirk} that switches over to the robust HLLE scheme in the vicinity of the shock in order to overcome the numerical instability problems of the Roe scheme.  Pandolfi and D'Ambrosio \cite{pand} have carried out more refined perturbation analysis by including the velocity perturbtions and  have observed that the schemes  in which the pressure perturbations damp out but create density perturbations that remain constant are strong carbulcle prone schemes. They have also observed that  schemes in which pressure perturbations remain constant  and do not interact with density perturbation, which also remain constant, are light carbuncle prone schemes, while schemes in which density perturbations damp out are carbuncle free schemes. Liou \cite{liou1} has conjectured  that the schemes that suffer  from numerical shock instability have a  pressure jump term in the numerical mass flux while those free from shock instability do not contain pressure jump term in the mass flux. Based on the conjecture of Liou, Kim et al. \cite{kim} have proposed a shock stable Roe scheme by reducing the contribution of the pressure jump terms in the mass flux. The findings of Quirk \cite{quirk} and Liou \cite{liou1} are qualitatively similar. 
Sanders et al. \cite{sand} have observed that  the instability is a result of inadequate cross-flow dissipation offered by the strictly upwind schemes and proposed a multi-dimensional, upwind dissipation to eliminate instability in the Roe scheme.
Dumbser et al. \cite{dumb-matrix} have developed a useful matrix stability method to evaluate stability of the numerical schemes and observed that the origin of the carbuncle phenomenon is localized in the upstream region of the shock.  Simon and Mandal  \cite{san1, san2, san3} have proposed strategies like anti-diffusion control (ADC) and Selective Wave-speed Modification (SWM) to cure the numerical instability problems in the HLLC and HLLEM Riemann solvers.  However, the cause of the numerical shock instabilities in the HLLC and HLLEM schemes is yet to be universally accepted. For example, Shen et al. \cite{shen}  and  Xie et al. \cite{xie} have identified the shear wave as the cause of the numerical shock instability in the HLLC and HLLEM  Riemann solvers, while Kemm \cite{kemm} has shown that resolution of the entropy or the contact wave in the HLLEM scheme also contributed to the numerical shock instabilities, but the contribution of the entropy wave is small as compared to the  contribution of the shear wave. Simon and Mandal \cite{san3} have showed that the HLLEMS scheme of Xie et al. is not free from the numerical shock instability problem and have demonstrated that the HLLEM scheme with anti-diffusion control (HLLEM-ADC) for both the entropy and shear waves is more effective in curing the numerical shock instabilities in the HLLEM scheme. 

Due to the numerical shock instability problems encountered by the HLLEM and HLLC schemes, several alternate approaches have been proposed to resolve the linearly degenerate discontinuities in the HLL scheme while  retaining its two wave structure. Linde \cite{linde} has shown that  the isolated discontinuities can be resolved in the HLL scheme by utilizing geometrical interpretation of the Rankine-Hugoniot jump conditions for scaling the state variables, while retaining the two-wave structure of the HLL scheme. In the HLL-CPS scheme proposed by Mandal and Panwar \cite{mandal}, the flux is split into convective and pressure parts and the pressure part of the flux is evaluated by the HLL method. The contact discontinuity is resolved by replacing the density jump terms of the HLL scheme by corresponding pressure jumps terms based on isentropic flow assumption. While the HLL-CPS scheme is capable of exact resolution of isolated contact discontinuity \cite{mandal}, the scheme has been shown to be incapable of accurate resolution of shear layers \cite{kita1}. In the HLL-BVD scheme proposed by Deng et al. \cite{deng}, the two-wave structure of the HLL scheme is retained and the resolution of contact discontinuity is achieved by reconstructing the density jump in the numerical dissipation terms with a jump-like function along with a boundary variation diminishing algorithm. The HLL-BVD  scheme \cite{deng} is capable of resolving  the isolated contact discontinuity exactly and is free from numerical shock instability problems like carbuncle, odd-even decoupling and kinked Mach stem.  The accuracy of the HLL-BVD scheme  is shown to exceed that of the HLLC scheme for certain test cases, but  the ability of the HLL-BVD scheme to resolve the shear wave is yet to be demonstrated.

Another major problem encountered by the approximate Riemann solvers, including the HLL scheme, is the admission of  nonphysical discrete results in low Mach number flow computations, especially on structured grids. Some of these solvers also encounter checkerboard-type pressure-velocity decoupling problem at very low Mach numbers. Therefore,  a number of different approaches like preconditioning \cite{guillard1, park-pre},  eigenvalue modification \cite{li-roe}, velocity jump scaling \cite{rieper1, rieper2, ossw}, velocity reconstruction \cite{th1},  anti-diffusion terms \cite{della2}, momentum jump scaling \cite{li-gu-hll} etc.,  have been taken up to improve performance  of these schemes in low Mach number flows. It has been shown by Guillard and Viozat \cite{guillard1} and Rieper \cite{rieper1, rieper2} that, in the limit as Mach number tends to zero, the pressure fluctuations for the discrete Euler equations obtained from the Roe scheme are of the order of Mach number, whereas  the pressure fluctuations  for the continuous Euler equations are of the order of Mach number squared. The non-physical behavior of the Roe scheme in low Mach number flows have been attributed to the wrong order of pressure  fluctuations in contrast with the continuous Euler equations. Guillard and Viozat have proposed a preconditioned Roe scheme \cite{guillard1} that has a correct pressure scaling with Mach number. Park et al. \cite{park-pre} have proposed a preconditioned HLLE+ scheme for low speed flows. Li and Gu \cite {li-roe} have observed that the preconditioned Roe schemes suffers from the global cutoff problem  and  have proposed an all speed Roe scheme by modifying the acoustic speed in the non-linear eigenvalues of the numerical dissipation terms \cite{li-roe}.  Rieper has proposed a low Mach number fix for the Roe scheme (LM-Roe) comprising of scaling of  the normal velocity jump in the numerical dissipation terms by a local Mach number function \cite{rieper2}. Using the low Mach number fix, Rieper has demonstrated accurate results for inviscid flow past a circular cylinder at very low Mach numbers with the modified Roe scheme. The low Mach number fix  of Rieper has been applied to both the normal and tangential velocity jumps in the numerical dissipation terms of the Roe scheme by Osswald et al. \cite{ossw}. Thornber et al. \cite{th1} have proposed a modification of the reconstructed velocities at the cell interface of the Godunov-type schemes like HLLC such that the velocity difference reduced with decreasing Mach number and the arithmetic mean is reached at zero Mach number. Thornber et al. have shown analytically and numerically that the numerical dissipation of the  scheme became constant in the  limit of zero Mach, while the numerical dissipation of the traditional schemes tended to infinity in the limit of zero Mach number.  Dellacherie \cite{della1} has observed that the Godunov-type schemes cannot be accurate at low Mach number due to loss of invariance. Dellacherie  has proposed central difference  to discretize the momentum flux or to discretize the  pressure gradient in momentum flux in order to recover the invariance property and improve accuracy in low Mach number flows. Dellacherie  et al. \cite{della2}  have also proposed  stable, all-Mach  Godunov-type schemes by adding  Mach number  based anti-diffusion  terms  to the Godunov-type schemes such that central-differencing is obtained in the limit of Mach zero. Yu et al. \cite{yu-hllem2} have implemented Dellacherie et al.'s low Mach correction in the HLLEM scheme. They have also proposed a scaling of the terms responsible for  the inaccurate behaviour of the HLLEM scheme at low Mach numbers with a Mach number function. 
 Li and Gu \cite{li-gu-hll} and Chen et al. \cite{chen-lm} have shown that the non-physical behavior of the HLL scheme in low Mach number flows can be overcome by suitable Mach number based scaling functions. Li and Gu \cite{li-gu-hll} have proposed a scaling  of the momentum jump terms in the momentum equations of the HLL scheme by a local Mach number function to overcome the non-physical behavior problem of the HLL scheme. Chen et al. \cite{chen-lm} have proposed low-diffusion Roe, HLL and Rusanov schemes by scaling all the velocity diffusion terms  with a local Mach number function. Li and Gu \cite{li-gu-hll} and Chen et al. \cite{chen-lm} have also demonstrated that the checkerboard pressure-velocity decoupling and global cut-off problems are avoided in their  improved HLL schemes.  Qu et al. \cite{qu-hllem} have  extended the HLLEMS scheme of Xie et al. \cite{xie}, into an all-speed scheme by scaling the momentum jumps with a local Mach number  function.
 
It is thus observed that considerable effort  is being made to extend the HLLEM and HLLC schemes  to all Mach numbers by addressing its shortcomings at very high and very low Mach numbers.  Several alternate formulations \cite{linde, mandal, deng}, different from HLLEM and HLLC, have also been proposed to introduce the linearly degenerate waves into the HLLE formulation while preserving shock stability at high Mach numbers.
In this paper, a simple and compact, all Mach number HLL-type scheme that is capable of resolving the contact and shear discontinuities and the low Mach flow features while preserving shock stability at high Mach numbers is proposed. A new method  for resolution of the shear layers and the  flow features at low Mach numbers based on a face-normal Mach number based velocity reconstruction method is proposed here. The contact wave is introduced in a manner similar to the HLLEM scheme. Shock stability at high Mach numbers is achieved by scaling with  face normal Mach number function with a pressure function.

The paper is organized into six sections. In Sections 2 and 3, the governing equations and the finite volume discretization are discussed. In Section 4, the flux evaluation in the proposed scheme is described. In Section 5, numerical results for several test cases are presented for speeds ranging from very low Mach number to very high Mach number. Finally, conclusions are presented in Section 6.

\section{Governing Equations}
The governing equations for the two-dimensional inviscid  compressible flow can be written  in their differential, conservative form as
\begin{equation} \label{2dge}
\dfrac{\partial{}\boldsymbol{\acute{U}}}{\partial{}t}+\dfrac{\partial{}\boldsymbol{\acute{F}(\acute{U})}}{\partial{}x}+\dfrac{\partial{}\boldsymbol{\acute{G}(\acute{U})}}{\partial{}y}=0
\end{equation}
where $\boldsymbol{\acute{U}}$, $\boldsymbol{\acute{F}(\acute{U})}$, $\boldsymbol{\acute{G}(\acute{U})}$ are the vectors of conserved variables, x-directional and y-directional fluxes respectively and can be written as
\begin{equation}
\boldsymbol{\acute{U}}=\left[\begin{array}{c} \rho \\ \rho{}u \\ \rho{}v \\ \rho{}E \end{array}\right] \hspace{1cm} \boldsymbol{\acute{F}(\acute{U})}=\left[\begin{array}{c} \rho{}u \\ \rho{}u^2+p \\ \rho{}uv \\ (\rho{}E+p)u \end{array}\right]  \hspace{1cm} \boldsymbol{\acute{G}(\acute{U})}=\left[\begin{array}{c} \rho{}v \\ \rho{}uv \\ \rho{}v^2+p \\ (\rho{}E+p)v \end{array}\right] 
\end{equation}
where $\rho$, u, v, p and E stand for density, x-directional and y-directional velocities in global coordinates, pressure and specific total energy. The system of equations is closed by the equation of state
\begin{equation}
p=(\gamma-1)\left(\rho{}E-\dfrac{1}{2}\rho(u^2+v^2)\right)
\end{equation} 
where $\gamma$ is the ratio of specific heats. In this paper, a calorifically perfect gas with $\gamma=1.4$ is considered.
The integral form  of the governing equations is given by
\begin{equation}\label{2dns}
	\frac{d}{dt}\int_{\Omega{}}\boldsymbol{\acute{U}}d\Omega{}+\oint_{\partial{}\Omega{}}\boldsymbol{(\acute{F},\acute{G}).n}dS=0
\end{equation}
where $\partial{}\Omega$ is the boundary of the control volume  $\Omega$ (area in case of two-dimensions) and  $\boldsymbol{n}$ denotes the outward pointing unit normal vector to the surface $\partial\Omega$.

\section{Finite Volume Discretization}
The finite volume discretization of the equation (4) for a structured, quadrilateral mesh can be written as
\begin{equation} \label{fv}
\dfrac{d\boldsymbol{\acute{U}}_i}{dt}=-\dfrac{1}{|\Omega|_i}\sum_{l=1}^4[\boldsymbol{(\acute{F},\acute{G})}_l.\boldsymbol{n}_l]\Delta{}s_l
\end{equation}
where $\boldsymbol{\acute{U}}_i$ is the cell averaged conserved state vector, $\boldsymbol{n}_l$ denotes unit interface normal vector and $\Delta{}s_l$ denotes the length of each interface. With the use of rotational invariance property of the Euler equations, the expression \ref{fv} can be written as 
\begin{equation} \label{fv}
\dfrac{d\boldsymbol{\acute{U}}_i}{dt}=-\dfrac{1}{|\Omega|_i}\sum_{l=1}^{4}(\mathcal{T}_{il})^{-1}\boldsymbol{F}(\mathcal{T}_{il}\boldsymbol{\acute{U}}_i,\mathcal{T}_{il}\boldsymbol{\acute{U}}_l)\Delta{}s_l
\end{equation}
where  $\boldsymbol{F}(\mathcal{T}_{il}\boldsymbol{\acute{U}}_i,\mathcal{T}_{il}\boldsymbol{\acute{U}}_l)$  is the   inviscid  face normal flux vector in locally rotated coordinate, $\mathcal{T}_{il}$ is the rotation matrix and ${\mathcal{T}_{il}}^{-1}$ is its inverse at the edge between cell $i$ and its neighbor $l$. 
 \begin{equation}
\mathcal{T}_{il}=\left(\begin{array}{ccccc}1& 0& 0& 0\\ 0& n_x& n_y& 0 \\  0& -n_y& n_x& 0 \\ 0& 0& 0& 1\end{array}\right) \hspace{1cm} and \hspace{1cm} {\mathcal{T}_{il}}^{-1}=\left(\begin{array}{ccccc}1& 0& 0& 0\\ 0& n_x&- n_y& 0 \\  0& n_y& n_x& 0 \\ 0& 0& 0& 1\end{array}\right)
\end{equation}
where $\boldsymbol{n}=(n_x,n_y)$ is the outward unit normal vector on the edge between cells $i$ and its neighbor $l$. 

\section{HLLE Scheme for all Mach number flows}
\subsection{Original HLL/HLLE  Flux }
The flux of the HLL  scheme in the direction normal to the cell interface can be written

\begin{equation} \label{flux-hll}
\boldsymbol{F}(\boldsymbol{U}_L,\boldsymbol{U}_R)=\dfrac{S_R\boldsymbol{F}_L-S_L\boldsymbol{F}_R}{S_R-S_L}+\dfrac{S_RS_L}{S_R-S_L}(\boldsymbol{U}_R-\boldsymbol{U}_L)
\end{equation} 
where $\boldsymbol{U}_L=\mathcal{T}_{il}\boldsymbol{\acute{U}}_l$ and $\boldsymbol{U}_R=\mathcal{T}_{il}\boldsymbol{\acute{U}}_i$ are the face normal state variables to the right and left of the interface, $\boldsymbol{F}_L=\boldsymbol{F}(\boldsymbol{U}_L)$ and $\boldsymbol{F}_R=\boldsymbol{F}(\boldsymbol{U}_R)$  are the face normal fluxes at the left and right sides of the interface,  $S_L$ and $S_R$ are the fastest left and right running wave speeds.

For the HLLE scheme,  the  wave speeds have been defined by Einfeldt \cite{einf1} as
\begin{equation} \label{wavespeed-hll}
S_L=min(0,u_{nL}-a_L,\tilde{u}_n-\tilde{a}) \hspace{1cm} S_R=max(0,u_{nR}+a_R,\tilde{u}_n+\tilde{a})
\end{equation}
where $u_{nL,R}$ and $a_{L,R}$ are the normal  velocities and sonic speeds to the left and right of the interface respectively, $\tilde{u}_n$ and $\tilde{a}$ are the Roe-averaged normal velocity and sonic speed at the interface. Einfeldt \cite{einf1} has showed that, with the above choice of wave speeds, the HLLE scheme  is entropy satisfying and positively conservative. 
\subsection{Simple extension  of HLLE Scheme for All Mach number Flows}
It can  be seen that several approaches have been followed to improve the ability of the HLL/HLLE scheme to resolve the contact and shear discontinuities.  In the approaches  proposed by Linde \cite{linde}, Mandal and Panwar \cite{mandal} and Deng et al. \cite{deng}, the two-wave structure of the original HLL scheme of Harten, Lax and van Leer \cite{hll} is  retained while a  three-wave structure is used in the approach of Toro et al. \cite{toro1}. In the present work, we do not add any wave structure for resolution of the shear layers  in the HLL framework. However, we try to achieve the resolution of the shear layers using a new low Mach correction method. This has an advantage in terms of numerical shock instability in contrast with the work of  Einfeldt et al. \cite{einf2} and Toro et al. \cite{toro1} where additional wave is added for resolving the shear layers.
\subsubsection {Resolution of Shear Wave and Low Mach Flow Features}
A modified version of the velocity reconstruction method proposed by Thornber et al. \cite{th1} for improving the accuracy of all Godunov-type scheme at low Mach numbers is utilized here for enhancing  the ability of the HLLE scheme  to resolve both the shear layers and the low Mach flow features. In the method proposed by Thornber et al. \cite{th1}, the velocity reconstruction is based on the local Mach number and is essentially aimed at improving the resolution of the low Mach flow features in the Godunov-type scheme like HLLC. However, we observe that the velocity reconstruction based on the local Mach number do not lead to exact resolution of the shear wave in the HLLE scheme, although it  helps in resolution of low Mach flow features. It has been found  that velocity reconstruction based on the face normal Mach number can lead to exact resolution of  the shear wave in the HLLE scheme. This is because of  the fact that the inability of the HLLE scheme to resolve the  the shear layers is primarily due to the presence of transverse velocity jump terms in the  transverse momentum flux equation. With the face normal Mach number based velocity reconstruction, the numerical dissipation due to the transverse velocity jump  in the transverse momentum flux  vanishes and hence the scheme  shall be able to resolve the shear layers efficiently. However, small disturbances have been observed in the vicinity of shock by Osswald et al. \cite{ossw} and Chen et al. \cite{chen-lm} when Mach number based function has been  used for scaling the velocity jumps. Hence, shock sensors have been used along with the Mach number based scaling functions for preserving shock structure. Therefore, in order to  resolve the low Mach flow features and the shear layers exactly in the HLLE scheme while preserving its shock capturing ability, we use the face normal Mach number along with a pressure function for reconstructing the velocities at the right and left of the interface.

The HLLE flux function  with  reconstructed velocities that is capable of resolving the shear wave and the flow features at low Mach numbers is written as
\begin{equation} \label{HLLE-TNP-flux}
\bold{F}(\bold{U}_L, \bold{U}_R)=\dfrac{S_R^*\bold{F}^*_L-S_L^*\bold{F}^*_R}{S_R^*-S_L^*}-\dfrac{S_R^*S_L^*}{2(S_R^*-S_L^*)}(\bold{U}^*_L-\bold{U}^*_R)
\end{equation}
where  $\boldsymbol{U}^*$ is the vector of reconstructed state variables to the right and left of the interface.
\begin{equation} \label{reconstructed-state-variables}
\boldsymbol{U}^*_{L/R}=\left[\begin{array}{c} \rho\\  \rho{}u^*_n \\ \rho{}u^*_t \\ \rho{}e^* \end{array}\right]_{L/R}=
\left[\begin{array}{c} \rho \\ \rho{}u^*_{n} \\ \rho{}u^*_{t}  \\  \dfrac{p}{\gamma-1}+\dfrac{1}{2}\rho(u^{*2}_n+u^{*2}_t) \end{array} \right]_{L/R}
\end{equation}
where  $u^*_n$ and $u^*_t$ are the reconstructed normal  and tangential velocities, $\bold{F}_L^*=\bold{F}(\bold{U}_L^*)$ and  $\bold{F}_R^*=\bold{F}(\bold{U}_R^*)$ are the vector of  fluxes to the left and right of the cell interface  based on the reconstructed state variables, $S_R^*$ and $S_L^*$ are the   wave speeds based on the reconstructed velocities  given by
\begin{equation}
S_R^*=max(0,u_{nR}^*+a_R, \tilde{u}_n+\tilde{a}) \hspace{1cm} S_L^*=min(0,u_{nL}^*-a_R, \tilde{u}_n-\tilde{a})
\end{equation}
The reconstructed velocities based on the face normal Mach number and a multi-dimensional pressure-based function are defined as
\begin{equation} \label{reconstructed-velocities}
u^*_L=\dfrac{1}{2}(u_L+u_R)+\dfrac{1}{2}z(u_L-u_R) \hspace{1cm} u^*_R=\dfrac{1}{2}(u_L+u_R)+\dfrac{1}{2}z(u_R-u_L)
\end{equation}
where \begin{equation} \label{z-znp}
z=1-(1-z_n)f_p
\end{equation}
\begin{equation} \label{fp}
f_p=min\left(\left(\dfrac{p_L}{p_R}\right), \left(\dfrac{p_R}{p_L}\right)\right)^3
\end{equation}
\begin{equation} \label{zn}
z_n=\left(1-min\left(max\left(\dfrac{u_{nL}}{a_L}, \dfrac{u_{nR}}{a_R}\right),1\right)\right)
\end{equation}
In order to detect the shock, the neighboring cells  must be considered. Therefore, the  pressure function $f_p$ at the interfaces $(i+\frac{1}{2}, j)$  and $(i, j+\frac{1}{2})$ is taken as 
\begin{equation}
\begin{array}{l}
fp_{i+\frac{1}{2},j}=min\left(fp_{i+\frac{1}{2},j},fp_{i,j+\frac{1}{2}},fp_{i+1,j+\frac{1}{2}},fp_{i,j-\frac{1}{2}},fp_{i+1,j-\frac{1}{2}}  \right) \\
fp_{i,j+\frac{1}{2}}=min\left(fp_{i,j+\frac{1}{2}},fp_{i-\frac{1}{2},j},fp_{i-\frac{1}{2},j+1},fp_{i+\frac{1}{2},j},fp_{i+\frac{1}{2},j+1}  \right)
\end{array}
\end{equation}
\subsubsection {Resolution of Contact Wave}
The contact wave is still not resolved the the HLLE flux function of equation (\ref{HLLE-TNP-flux}). The contact wave can be introduced into the HLLE scheme  by following either of the HLLEM \cite{einf2}, the HLL-CPS \cite{mandal} or the HLL-BVD \cite{deng} approaches. In the HLLEM approach, the contact wave is introduced by adding anti-diffusion terms corresponding to  the contact wave into the HLLE scheme. In the HLL-CPS scheme, the contact wave is introduced by replacing the density jumps  in the numerical dissipation terms with pressure jumps based on the isentropic assumption. In the HLL-BVD scheme, the contact wave is introduced by replacing the density jumps in the numerical dissipation terms with a jump-like function and by a boundary variation diminishing algorithm.

In the present work, the HLLEM approach of adding anti-diffusion terms  is followed for  introducing the contact wave into the HLLE scheme. Thus the HLLE flux function for resolving the  shear wave, the contact wave and the low Mach flow features can be written as 
\begin{equation}\label{HLLE-TNP-flux}
\bold{F}(\bold{U}_L, \bold{U}_R)=\dfrac{S_R^*\bold{F}^*_L-S_L^*\bold{F}^*_R}{S_R^*-S_L^*}-\dfrac{S_R^*S_L^*}{2(S_R^*-S_L^*)}(\bold{U}^*_L-\bold{U}^*_R-\delta_2\alpha_2\bold{\bar{R}_2})
\end{equation}
where $\delta_2$ is the anti-diffusion coefficient, $\alpha_2=\Delta{}\rho-\dfrac{\Delta{}p}{\bar{a}^2}$ is the wave-strength and $\bold{\bar{R}}_2=\left[ 1, \\ \bar{u}_n, \\ \bar{u}_t, \\ \frac{1}{2}(\bar{u}_n^2+\bar{u}_t^2)\right]^T$  is  the right eigenvector of the contact wave in the flux Jacobian matrix, and $\Delta=(.)_L-(.)_R)$. 
Here, the arithmetic averaged values (denoted by superscript $(\bar{})$ ) are used instead of the Roe-averaged values for the right eigenvector terms and it  will be shown later  that  the arithmetic averaged values lead to exact resolution of isolated contact discontinuity. The anti-diffusion coefficient proposed by Park and Kwon \cite{park} and commonly used  in the HLLEM scheme is $\delta_2=\dfrac{\tilde{a}}{\tilde{a}+|\tilde{u}_n|}$. However, this anti-diffusion coefficient may lead to numerical shock instabilities at high Mach numbers. Therefore, in the present scheme, the  anti-diffusion coefficient for the contact wave that is capable of suppressing the numerical shock instabilities and resolving the contact discontinuity is defined as
\begin{equation} \label{anti-diffusion-coefficient}
\delta_2=1-z
\end{equation} 
where $z$ is the face normal Mach number and  pressure  based function defined previously in equation (\ref{z-znp}).

The proposed scheme shall be referred to as the HLLE-TNP scheme, that is, the HLLE scheme with $\bold{T}$hornber-type velocity reconstruction based on face $\bold{N}$ormal velocity and $\bold{P}$ressure  function.

In the next section, we present proof of the exact isolated contact and shear discontinuity resolving ability of the proposed HLLE-TNP scheme.
\subsubsection {Proof of Exact Resolution of Isolated Contact Discontinuities by HLLE-TNP}
Across the isolated  contact and shear discontinuities, the face normal Mach numbers are  zero ($u_{nL}=u_{nR}=0$) while the pressures are constant ( $p_L=p_R=p$).  Therefore, the velocity reconstruction function shown in equation (\ref{z-znp}) reduces to

\begin{equation}
z=1-\left[1-\left(1-min\left(max\left(\dfrac{u_{nL}}{a_L}, \dfrac{u_{nR}}{a_R}\right),1\right)\right)\right]\left(\dfrac{p}{p}\right)^3=0
\end{equation}
The anti-diffusion coefficient for the contact wave shown in equation (\ref{anti-diffusion-coefficient}) becomes $\delta_2=1-z=1$.
The reconstructed velocities to the left and right of the interface shown in  equation (\ref{reconstructed-velocities}) becomes 
\begin{equation}
\begin{array}{l} 
u^*_{nL}=\dfrac{1}{2}(u_{nL}+u_{nR})+\dfrac{1}{2}z(u_{nL}-u_{nR})=0 \\
u^*_{nR}=\dfrac{1}{2}(u_{nL}+u_{nR})+\dfrac{1}{2}z(u_{nR}-u_{nL})=0 \\
u^*_{tL}=\dfrac{1}{2}(u_{tL}+u_{tR})+\dfrac{1}{2}z(u_{tL}-u_{tR})= \dfrac{1}{2}(u_{tL}+u_{tR})=\bar{u}^*_t\\
u^*_{tR}=\dfrac{1}{2}(u_{tL}+u_{tR})+\dfrac{1}{2}z(u_{tR}-u_{tL})= \dfrac{1}{2}(u_{tL}+u_{tR})=\bar{u}^*_t
\end{array}
\end{equation}
where superscript $(\bar{})$ denote  the arithmetic averaged values of the variables to the right and left of the interface. The reconstructed state variables shown in equation (\ref{reconstructed-state-variables}) and the fluxes at the right and left of the interface become
\begin{equation} 
\boldsymbol{U}^*_{L}=\left[\begin{array}{c} \rho_L \\ 0 \\ \rho_L\bar{u}^*_{t}  \\  \dfrac{p_L}{\gamma-1}+\dfrac{1}{2}\rho_L\bar{u}^{*2}_t \end{array} \right] \hspace{1cm} \boldsymbol{U}^*_{R}=\left[\begin{array}{c} \rho_R \\ 0 \\ \rho_R\bar{u}^*_{t} \\  \dfrac{p_R}{\gamma-1}+\dfrac{1}{2}\rho_R\bar{u}^{*2}_t \end{array} \right] \hspace {1cm} \boldsymbol{F}^*_{L/R}=\left[\begin{array}{c} 0 \\ p \\ 0 \\ 0 \end{array}\right]
\end{equation}

The flux function at the cell interface shown in equation (\ref{HLLE-TNP-flux}) becomes
\begin{equation}
\begin{split}
&\bold{F}(\bold{U}_L, \bold{U}_R)=\dfrac{1}{S_R^*-S_L^*}\left[S_R^*\left(\begin{array}{c} 0 \\ p \\ 0 \\ 0 \end{array}\right)-S_L^*\left(\begin{array}{c} 0 \\ p \\ 0 \\ 0 \end{array}\right)\right]-  \\ &
\dfrac{S_R^*S_L^*}{2(S_R^*-S_L^*)}\left[ \begin{array}{c}\rho_L-\rho_R-(\rho_L-\rho_R-\dfrac{p_L}{\bar{a}^2}+\dfrac{p_R}{\bar{a}^2}) \\ 0-0-0\left(\rho_L-\rho_R-\dfrac{p_L}{\bar{a}^2}+\dfrac{p_R}{\bar{a}^2}\right) \\ \rho_L\bar{u}^*_t-\rho_R\bar{u}^*_t-\bar{u}^*_t\left(\rho_L-\rho_R-\dfrac{p_L}{\bar{a}^2}+\dfrac{p_R}{\bar{a}^2}\right) \\ \dfrac{p_L-p_R}{\gamma-1}+\dfrac{1}{2}(\rho{}_L\bar{u}^{*2}_t-\rho{}_R\bar{u}^{*2}_t)-\dfrac{\bar{u}^{*2}_t}{2}\left(\rho_L-\rho_R-\dfrac{p_L}{\bar{a}^2}+\dfrac{p_R}{\bar{a}^2}\right)\end{array}\right]
\end{split}
\end{equation}

All the density and pressure jump terms cancel when arithmetic averaged values are used for the anti-diffusion terms  of the contact wave, which will not be the case with Roe averaged values. Therefore, the flux function at the interface reduces to
\begin{equation}
\boldsymbol{F}(\boldsymbol{U}_L, \boldsymbol{U}_R)=\left[\begin{array}{c} 0 \\ p \\ 0 \\ 0 \end{array}\right]
\end{equation}
Thus, the flux function comprise of  the physical flux only with zero numerical dissipation. Hence, the proposed scheme  shall be capable of resolving the isolated contact and shear discontinuities exactly. 

In the following sections, the suitability of the proposed scheme for low Mach number flows is demonstrated through asymptotic analysis while the stability of the proposed scheme is demonstrated through  linear perturbation and matrix stability analyses.
\subsection{Asymptotic Analysis}
The asymptotic analysis of these schemes are carried out  in a manner similar to Guillard and Viozat \cite{guillard1} and Rieper \cite{rieper1, rieper2}.  
In the asymptotic analysis,  a cartesian grid of uniform mesh size is considered  where $\bold{i}=(i,j)$ is the index of cell and $\upsilon(\bold{i})=((i+1,j), (i-1,j), (i,j+1),(i,j-1))$ are the neighbors of the cell $\bold{i}$.
For low speed flows the wave speeds can be approximated as $S_R=u_n+a$ and $S_L=u_n-a$. Therefore, the following simplifications in equation(\ref{flux-hll}) can be made :
\begin{equation}
\dfrac{S_RS_L}{(S_R-S_L)}=\dfrac{u_n^2-a^2}{2a}, \hspace{1cm} \dfrac{S_R+S_L}{S_R-S_L}=\dfrac{u_n}{a}
\end{equation}
The variables are non-dimensionalized as follows
\begin{equation}
\bar{\rho}=\dfrac{\rho}{\rho{}_*} \hspace{1cm} \bar{u}=\dfrac{u}{u_*} \hspace{1cm} \bar{p}=\dfrac{p}{\rho_*a_*^2} \hspace{1cm} \bar{e}=\dfrac{e}{a_*^2}\hspace{1cm}  \bar{t}=\dfrac{tu_*}{S_*} \hspace{1cm} \bar{\Omega}=\dfrac{\Omega}{S_*^2}
\end{equation}
where $\rho_*$ is reference density, $u_*$ is reference velocity, $a_*$ is reference speed of sound, $S_*$ is reference length. 
The face normal Mach number function in non-dimensional form is 
\begin{equation}
z_n=\dfrac{u_n}{a}=\dfrac{u_n}{u_*}\dfrac{a_*}{a}\dfrac{u_*}{a_*}=\dfrac{\bar{u}_n}{\bar{a}}M_*=\bar{z}_nM_*
\end{equation}
It can be seen that the face normal Mach number function is of the order of Mach number, ie, $\mathcal{O}(M)$. At low Mach numbers,  the pressure function shown in equation (\ref{fp}) approaches unity and hence the velocity reconstruction function $z$ shown in equation (\ref{z-znp})  reduces to  $\bar{z}=\bar{z}_n$.
The HLLE-TNP scheme shown in (\ref{fv}, \ref{flux-hll}) can be written in non-dimensional form, after rotation,  as 
\begin{equation}
\begin{split}
& |\bar{\Omega}|\dfrac{\partial{}\bar{U}_i}{\partial{}\bar{t}}+  \sum_{\bold{l}\epsilon\upsilon(\bold{i})}\left[\begin{array}{c} \bar{\rho}\bar{u}_n \\ \bar{\rho}\bar{u}\bar{u}_n+\dfrac{\bar{p}}{M_*^2}n_x\\  \bar{\rho}\bar{v}\bar{u}_n+\dfrac{\bar{p}}{M_*^2}n_y \\ (\bar{\rho}\bar{e}+\bar{p})\bar{u_n}\end{array}\right]_{il}+ \left(\dfrac{\bar{u}_nM_*}{2\bar{a}}\right)_{il}\left[\begin{array}{c} \Delta_{il}(\bar{\rho}\bar{u}^*_n) \\ \Delta_{il}(\bar{\rho}\bar{u}^*\bar{u}^*_n)+\left(\dfrac{n_x}{M_*^2}\right)_{il}\Delta_{il}(\bar{p})\\ \Delta_{il}(\bar{\rho}\bar{v}^*\bar{u}^*_n)+ \left(\dfrac{n_y}{M_*^2}\right)_{il} \Delta_{il}(\bar{p})\\ \Delta_{il}(\bar{p}\bar{u}_n)\end{array}\right]\Delta{}s_{il}-  \\ 
&  \sum_{\bold{l}\epsilon\upsilon(\bold{i})}\left(\dfrac{\bar{u}_n^2M_*}{2\bar{a}}-\dfrac{\bar{a}}{2M_*}\right)_{il}\left[\begin{array}{c} \Delta_{il}(\bar{\rho})-(1-\bar{z}M_*)(\Delta{}_{il}\bar{\rho}-\dfrac{\Delta_{il}\bar{p}}{\bar{a}^2_{0,il}}) \\ \Delta_{il}(\bar{\rho}\bar{u}^*)-(1-\bar{z}M_*)(\Delta{}_{il}\bar{\rho}-\dfrac{\Delta_{il}\bar{p}}{\bar{a}^2_{0,il}})\bar{u}_{il}\\ \Delta_{il}(\bar{\rho}\bar{v}^*)-(1-\bar{z}M_*)(\Delta{}_{il}\bar{\rho}-\dfrac{\Delta_{il}\bar{p}}{\bar{a}^2_{0,il}})\bar{v}_{il} \\ \Delta_{il}(\bar{\rho}\bar{e}^*)-(1-\bar{z}M_*)(\Delta{}_{il}\bar{\rho}-\dfrac{\Delta_{il}\bar{p}}{\bar{a}^2_{0,il}})\dfrac{\bar{q}^2_{il}}{2}\end{array}\right]\Delta{}s_{il}=0
\end{split}
\end{equation}
Expanding the terms asymptotically using equation (\ref{expansion}) and arranging the terms in equal power of $M_*$, we get 

1. Order of $M_*^0$
\begin{itemize}
\item{Continuity Equation}
\begin{equation}
|\bar{\Omega}|\dfrac{\partial\bar{\rho}_{0i}}{\partial{}\bar{t}}+\sum_{\bold{l}\epsilon\upsilon(\bold{i})}\left((\bar{\rho}_0\bar{u}_{n0})_{il}+\dfrac{\bar{a}_{0,il}}{2}\left(\bar{z}_{il}(\Delta_{il}\bar{\rho}_0-\dfrac{\Delta_{il}\bar{p}_0}{\bar{a}^2_{0,il}})+\dfrac{\Delta_{il}p_1}{\bar{a}^2_{0,il}}\right)\right)\Delta{}s_{il}=0
\end{equation}
\item{x-momentum equation}
\begin{equation}
\begin{split}
& |\bar{\Omega}|\dfrac{\partial(\bar{\rho}_0\bar{u}_0)_i}{\partial{}\bar{t}}+\sum_{\bold{l}\epsilon\upsilon(\bold{i})}\left((\bar{u}_{n0}\bar{\rho}_{0}\bar{u}_0)_{il}+(\bar{p}_2n_x)_{il}+\left(\dfrac{\bar{u}_{n0}}{2\bar{a}_{0,il}}n_x\right)_{il}\Delta_{il}\bar{p}_1\right)\Delta{}s_{il} \\ &
+\sum_{\bold{l}\epsilon\upsilon(\bold{i})}\dfrac{\bar{a}_{0,il}}{2}\left(\bar{z}_{il}\Delta_{il}(\bar{\rho}_0\bar{u}_0)-\left(\bar{z}\dfrac{\bar{u}_0}{\bar{a}_0^2}\right)_{il}\Delta{}_{il}\bar{p}_0+\left(\dfrac{\bar{u}_0}{\bar{a}_0^2}\right)_{il}\Delta{}_{il}\bar{p}_1\right)\Delta{}s_{il}=0
\end{split}
\end{equation}
\item{y-momentum equation}
\begin{equation}
\begin{split}
& |\bar{\Omega}|\dfrac{\partial(\bar{\rho}_0\bar{v}_0)_i}{\partial{}\bar{t}}+\sum_{\bold{l}\epsilon\upsilon(\bold{i})}\left((\bar{u}_{n0}\bar{\rho}_{0}\bar{v}_0)_{il}+(\bar{p}_2n_y)_{il}+\left(\dfrac{\bar{u}_{n0}}{2\bar{a}_0}n_y\right)_{il}\Delta_{il}\bar{p}_1\right)\Delta{}s_{il} \\ &
+\sum_{\bold{l}\epsilon\upsilon(\bold{i})}\dfrac{\bar{a}_{0,il}}{2}\left(\bar{z}_{il}\Delta_{il}(\bar{\rho}_0\bar{v}_0)-\left(\bar{z}\dfrac{\bar{v}_0}{\bar{a}_0^2}\right)_{il}\Delta{}_{il}\bar{p}_0+\left(\dfrac{\bar{v}_0}{\bar{a}_0^2}\right)_{il}\Delta{}_{il}\bar{p}_1\right)\Delta{}s_{il}=0
\end{split}
\end{equation}
\end{itemize}

2. Order of $M_*^{-1}$ 
\begin{itemize}
\item{Continuity Equation}
\begin{equation}
\sum_{\bold{l}\epsilon\upsilon(\bold{i})}\dfrac{1}{2a_{0,il}}\Delta_{il}\bar{p}_0\Delta{}s_{il}=0
\end{equation}
Here $\sum_{\bold{l}\epsilon\upsilon(\bold{i})}\Delta_{il}\bar{p}_0=0$ since the length of the interface $\Delta{}s_{il}$ is positive and hence $\bar{p}_0=cte\forall{}\bold{i}$.
\item{x-momentum equation}
\begin{equation}
 \sum_{\bold{l}\epsilon\upsilon(\bold{i})}\left((\bar{p}_1n_x)_{il}+\left(\dfrac{\bar{u}_{n0}}{2\bar{a}_0}n_x\right)_{il}\Delta_{il}\bar{p}_0-\left(\dfrac{\bar{a}_0}{2}\dfrac{\bar{u}_0}{\bar{a}_0^2}\right)_{il}\Delta_{il}\bar{p}_0\right)\Delta{}s_{il}=0
\end{equation}
\item{y-momentum equation}
\begin{equation}
 \sum_{\bold{l}\epsilon\upsilon(\bold{i})}\left((\bar{p}_1n_y)_{il}+\left(\dfrac{\bar{u}_{n0}}{2\bar{a}_0}n_y\right)_{il}\Delta_{il}\bar{p}_0-\left(\dfrac{\bar{a}_0}{2}\dfrac{\bar{v}_0}{\bar{a}_0^2}\right)_{il}\Delta_{il}\bar{p}_0\right)\Delta{}s_{il}=0
\end{equation}
Since  $\sum_{\bold{l}\epsilon\upsilon(\bold{i})}\Delta_{il}\bar{p}_0=0$ is obtained from the continuity equation, the x- and y-momentum equation reduces to 
\begin{equation} \label{p-mom}
 \sum_{\bold{l}\epsilon\upsilon(\bold{i})}\bar{p}_1n_x)_{il}=0 \hspace{2cm}
 \sum_{\bold{l}\epsilon\upsilon(\bold{i})}\bar{p}_1n_y)_{il}=0
\end{equation}
Equation (\ref{p-mom}) implies that 
\begin{equation}
\bar{p}_{1,i-1,j}-\bar{p}_{1,i+1,j}=0, \;\bar{p}_{1,i,j-1}-\bar{p}_{1,i,j+1}=0, 
\end{equation}
This implies that $\bar{p}_1=cte\forall{}\bold{i}$ where $cte$ is a constant. Hence the  HLLE-TNP scheme for the discrete Euler equations supports pressure fluctuation of the type $p(x,t)=P_0(t)+M_*^2p_2(x,t)$ as in the case of the continuous Euler equations at low Mach numbers. 

3. Order of $M_*^{-2}$
\begin{equation}
\bar{p}_{0,i-1,j}-\bar{p}_{1,i+1,j}=0, \;\bar{p}_{0,i,j-1}-\bar{p}_{1,i,j+1}=0, 
\end{equation}
This implies that $\bar{p}_0=cte\forall{}\bold{i}$ where $cte$ is a constant.
\end{itemize}

It can be seen that the pressure fluctuations in the proposed HLLE-TNP schemes are of the type represented by the equation $p(x,t)=P_0(t)+M_*^2p_2(x,t)$. Hence, the proposed scheme shall be capable of resolving the  low Mach flow features.  Asymptotic analysis of the  classical HLLE scheme is placed in Appendix I. It can be seen that the pressure fluctuations are of the type  $p(x,t)=P_0(t)+M_*p_1(x,t)$.

\subsection{Linear Perturbation Analysis}
Linear Perturbation analysis of the numerical scheme is helpful  for determining the stability of the scheme for high speed flow problems. The linear perturbation analysis of the upwind schemes have been pioneered  by Quirk \cite{quirk} and refined by Pandolfi and  D'Ambrosio \cite{pand}. The evolution of the density, shear velocity and pressure perturbations in  the proposed schemes is carried out in a manner similar to Quirk \cite{quirk} and Pandolfi \cite{pand}. The normalized flow properties are described as follows
\begin{equation}
\rho=1\pm \hat{\rho}, \hspace{1cm} u=u_0\pm \hat{u}, \hspace{1 cm} v=0, \hspace{1cm}  p=1\pm\hat{p}
\end{equation}
where $\hat{\rho}, \hat{u}$ and $\hat{p}$ are the density, shear velocity and pressure perturbations. It may be noted that, here $u$ is the shear velocity and $v$ is the normal velocity.
The evolution of the density, shear velocity and pressure perturbation for the HLLE, Roe, HLLC, HLLEM and the proposed  schemes are shown in Table \ref{pert-anal-results}. It can be seen from the  table that the density, shear velocity and pressure perturbations are damped in the HLLE scheme. In the Roe, HLLC and HLLEM schemes, the density perturbations are fed by the pressure perturbations while the shear velocity perturbations remain unchanged. In the proposed HLLE-TNP scheme, the  pressure switch $f_p$ and the density and pressure perturbations influence the evolution of the density perturbations. Across a strong shock, the value of $f_p$ becomes close to zero and hence the density and shear velocity perturbations are damped in a manner  similar to the HLLE scheme. Since the density and shear velocity perturbations are damped in presence of  strong shock in the proposed HLLE-TNP scheme, the scheme is likely to be free from numerical shock instabilities. 
\begin{table}[H]
\caption{Result of Linear Perturbation Analysis of HLL-type scheme}
\label{pert-anal-results}
\begin{doublespace}
\begin{tabular}{|l|l|l|l|l|}
\hline Serial No & Scheme & $\hat{\rho}^{n+1}=$ & $\hat{u}^{n+1}=$ & $\hat{p}^{n+1}=$\\ 
\hline  1 & HLLE & $\hat{\rho}^n(1-2\nu)$ & $\hat{u}^n(1-2\nu)$ & $\hat{p}^n(1-2\nu)$\\ 
\hline  2 & Roe/HLLEM/HLLC & $\hat{\rho}^n-\dfrac{2\nu}{\gamma}\hat{p}^n$ & $\hat{u}^n$ & $\hat{p}^n(1-2\nu)$\\   
\hline  3 & HLLE-TNP& $\hat{\rho}^n(1-2\nu{}(1-f_p))-2\nu{}\dfrac{f_p}{\gamma}\hat{p}^n$ & $\hat{u}^n(1-2\nu(1-f_p))$ & $\hat{p}^n(1-2\nu)$\\ 
\hline
\end{tabular}
\end{doublespace}
\end{table}
\subsection{Matrix Stability Analysis}
In the present work, matrix stability analysis  is carried out in a manner similar to Dumbser et al. \cite{dumb-matrix}. A 2D computational domain $[0,1] \times [0,1]$ is considered and the domain is discretized into $11 \times 11$ grid points.  The  raw state is a steady normal shock wave with some perturbations which can be expressed as
\begin{equation}
\boldsymbol{U}_\bold{i}=\boldsymbol{U}^0_\bold{i}+\delta{}\boldsymbol{U}_\bold{i}
\end{equation}
where $\bold{i}=(i,j)$ is the index of cell, $\boldsymbol{U}^0_\bold{i}$ is the solution of a steady shock and $\delta{}\boldsymbol{U}_\bold{i}$ is a small numerical random perturbation.
Substituting the expression into finite volume formulation, and after some evolution, the perturbation can be expressed as
\begin{equation}
\left(\begin{array}{c} \delta{}\boldsymbol{U}_1 \\ . \\ . \\ . \\ \delta{}\boldsymbol{U}_M\end{array}\right)=exp^{St}\left(\begin{array}{c} \delta{}\boldsymbol{U}_1 \\ . \\ . \\ . \\ \delta{}\boldsymbol{U}_M\end{array}\right)_0
\end{equation}
where S is  the stability matrix based on the Riemann solver. The perturbations will remain bounded if the maximum of the real part of the eigenvalues of S is non-positive, i.e.,
\begin{equation}
max(Re(\lambda(S)))\le 0
\end{equation}
The initial data is provided by the exact Rankine-Hugoniot solution in x-direction. The upstream and downstream states  of the primitive variables are
\begin{equation}
\begin{array}{l}
W_L=(\rho, u, v, p)_L=\left(1, 1, 0, \dfrac{1}{\gamma{}M^2_a}\right), x< 0.5 \\ 
W_R=(\rho, u, v, p)_R=\left(f(M_a),\dfrac{1}{f(M_a)}, 0, \dfrac{g(M_a)}{\gamma{}M^2_a}\right), x> 0.5
\end{array}
\end{equation}
where $M_a$ is the upstream Mach number,
$f(M_a)=\left( \dfrac{2}{\gamma+1}\dfrac{1}{M^2_a}+\dfrac{\gamma-1}{\gamma+1}\right)^{-1}$,  
$g(M_a)=\left( \dfrac{2\gamma}{\gamma+1}{M^2_a}-\dfrac{\gamma-1}{\gamma+1}\right)$.  

\begin{figure}[H]
	\begin{center}
	\includegraphics[width=225pt]{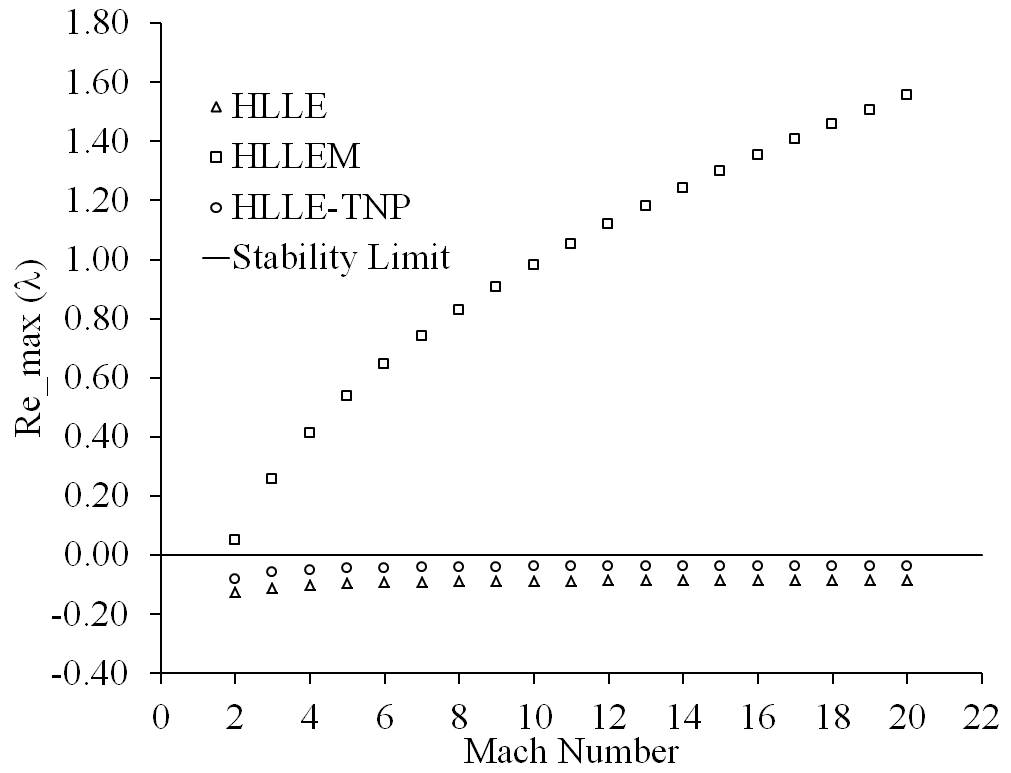} 
	\caption{Plot of Maximum Real Eigenvalue of $S$ vs upstream Mach Number for the  proposed HLLE-TNP  scheme. The results of the HLLE  and HLLEM schemes are also shown for comparison}
	\label{max-eigenvalue}
	\end{center}
\end{figure}

\begin{figure}[H]
	\begin{center}
	\includegraphics[width=225pt]{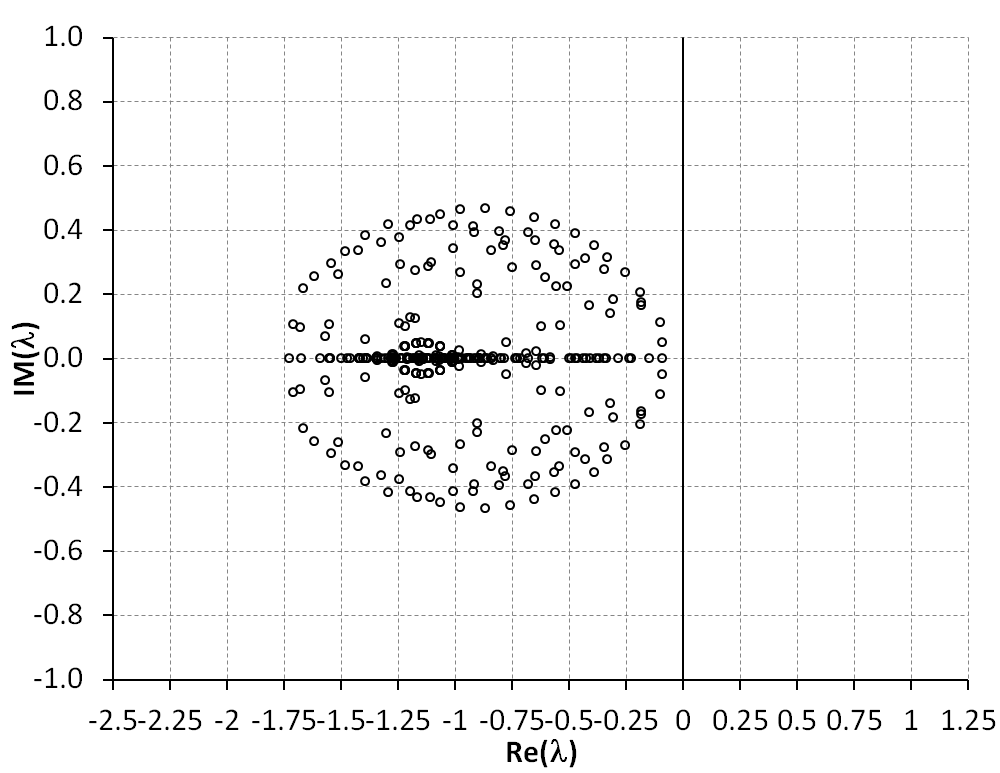} 	\includegraphics[width=225pt]{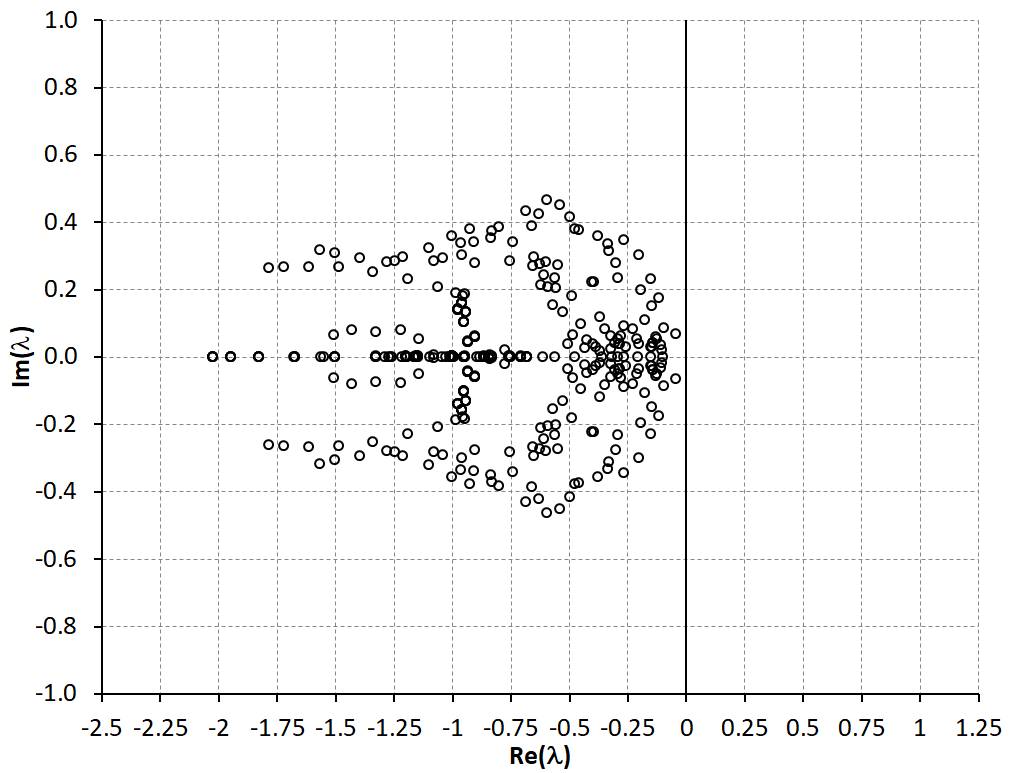} 
	\caption{Distribution of the eigenvalues  of $S$ in the complex plane for the  HLLE and proposed HLLE-TNP schemes. The results are shown for a grid size of $11\times11$  and an upstream  Mach number M=7.0.}
	\label{eigenvalue-hlle}
	\end{center}
\end{figure}

A random perturbation of $10^{-6}$ is introduced to all the conserved variables  of all the grid cells.  The matrix stability analysis can be carried out with either a `thin shock' comprising of  only upstream and downstream  values or with a `shock structure' comprising of an intermediate state. In the present work, the analysis is carried out for a `thin' shock. The plot of the maximum real eigenvalues  with respect to upstream Mach number  for the HLLE scheme, HLLEM scheme and proposed HLLE-TNP scheme are  shown in Fig. \ref{max-eigenvalue}. It can be seen from the figure that  the maximum  real eigenvalue of the HLLEM schemes is positive for all the Mach number considered in the analysis. Thus, the HLLEM scheme is unstable.  The maximum real eigenvalues of the proposed HLLE-TNP scheme are negative  for all the Mach numbers, just like the  classical HLLE scheme. Therefore, the proposed HLLE-TNP scheme shall be free from numerical shock instabilities like the HLLE scheme. The  eigenvalue distribution of  the matrix S in the complex plane for the HLLE and proposed HLLE-TNP scheme  for a representative inflow Mach number of 7.0 are shown in Fig. \ref{eigenvalue-hlle}. The  eigenvalue plot of the HLLE scheme is almost identical to the result obtained by Dumbser et al. \cite{dumb-matrix}. It can be seen that all the real eigenvalues of the HLLE and the proposed HLLE-TNP scheme are negative. 

In the next section, results are presented for a number of test cases covering very high to very low Maach numbers in order to demonstrate the robustness and accuracy of the proposed scheme. 

\section{Results}
The performance of the proposed HLLE-TNP  scheme  is  evaluated by solving the problems involving contact and shear discontinuity, monotonicity, planar shock, double Mach reflection, forward facing step, blunt body, expansion corner, low speed cylinder and flat plate boundary layer. Results of the HLLE amd HLLEM schemes for these test cases are also shown for comparison. First order results are shown for  the inviscid, high speed test cases since the numerical shock instabilities are prominently  visible in first order schemes. However, results are generated using second order schemes for the viscous test cases.
\subsection{Test for moving contact discontinuity and monotonicity}
\begin{figure}[H]
\begin{center}
\includegraphics[width=200pt]{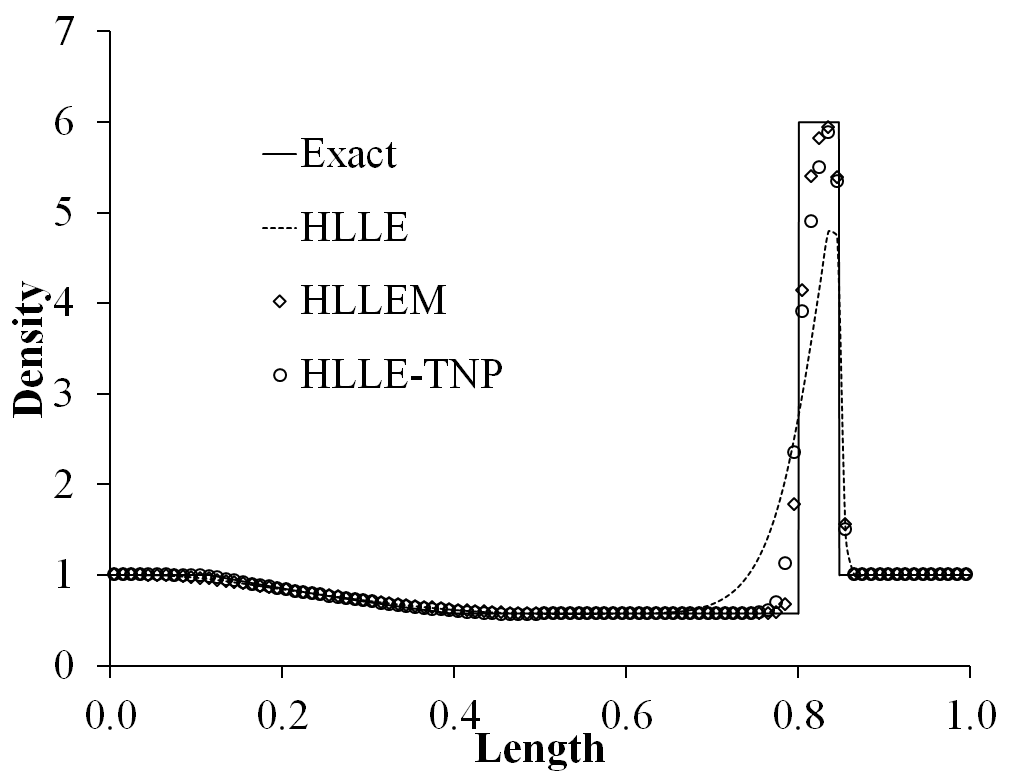}\includegraphics[width=200pt]{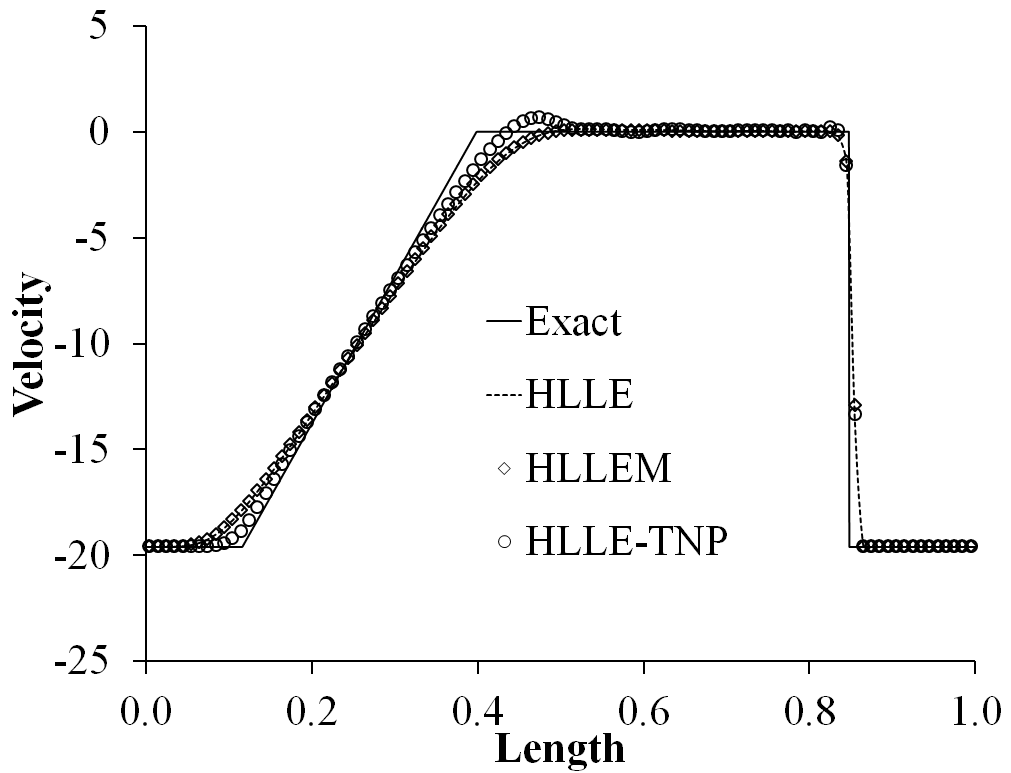}
\includegraphics[width=200pt]{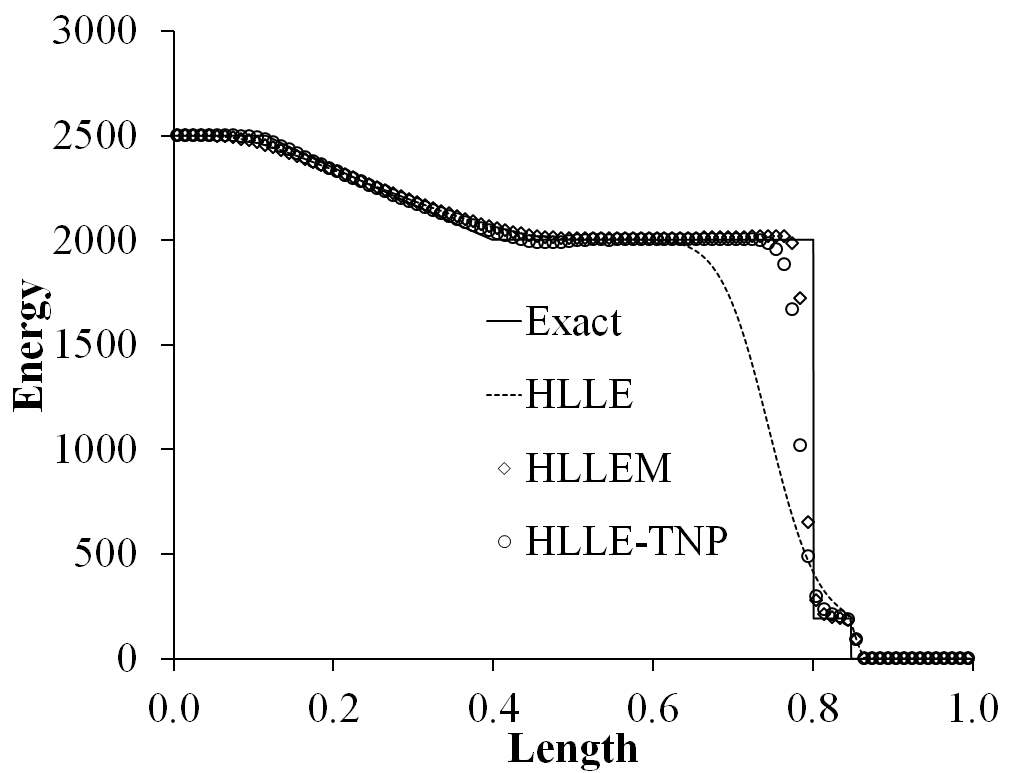}\includegraphics[width=200pt]{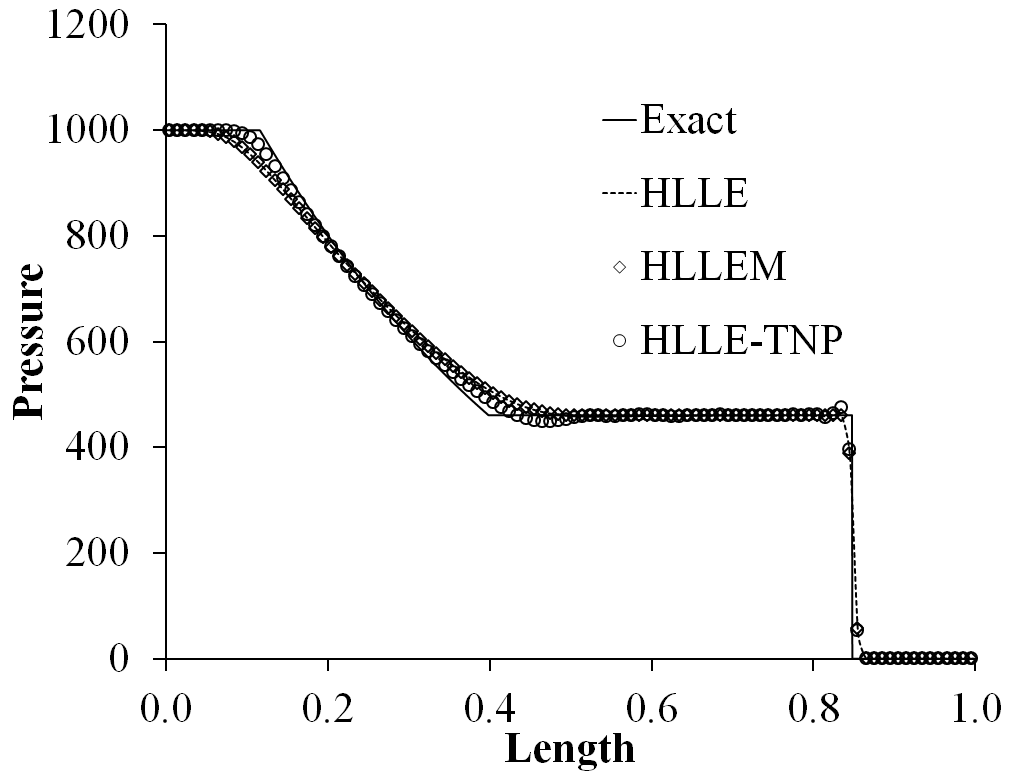}
\caption{Results of the HLLE , HLLEM and the proposed HLLE-TNP schemes for test case with $x_0=0.8$. The numerical and exact solutions are compared at time t=0.012.}
\label{hlle-test-case5}
\end{center}
\end{figure}

This test case has been proposed by Toro \cite{toro-book} and has been designed to assess the ability of the numerical methods to resolve slowly moving contact discontinuities. This test is also used to demonstrate the monotone behavior of the numerical method. The test has a strong shock running rightwards, a stationary contact and a left running expansion wave. The computational domain length is taken as unity and the domain is discretized into 100 cells. A CFL number of 0.9 is used. The values of the left state are  $(\rho,u,p)_L$ = (1.0, -19.59745, 1000.0) while the values of the right state are $(\rho,u,p)_R$= (1.0, -19.59745, 0.01). The left and right states are separated at  x$_{0}=0.8$.  The results are shown in  Fig. \ref{hlle-test-case5}. The HLLE-TNP scheme with velocity reconstruction based on Mach number and pressure function is able to generate monotone results around the shock shock without overshoot in density, velocity and pressure. The moving contact discontinuity, represented by the jump in density and specific internal energy, is smeared in the classical HLLE scheme while the HLLEM and proposed HLLE-TNP schemes  are able to resolve the moving contact discontinuity efficiently with 4-5 cells. 
\subsection{Two-dimensional inviscid shear flow}
\begin{figure}[H]
\begin{center}
\includegraphics[width=200pt]{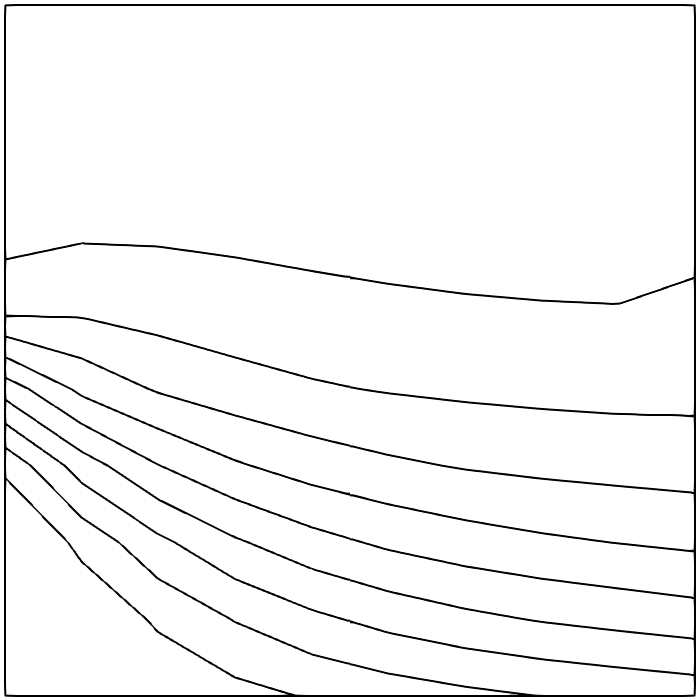} \hspace{1 cm} \includegraphics[width=200pt]{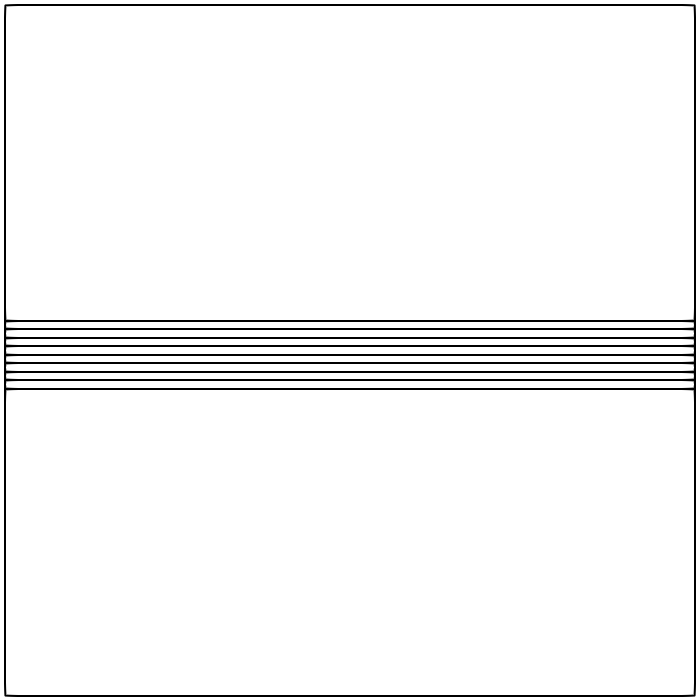} \\
(a) HLLE Scheme \hspace{5cm} (b) HLLE-TNP Scheme
\caption{Density contours computed by the HLLE and HLLE-TNP schemes for inviscid shear flow. The results are shown after 1000 iterations}
\label{shear-flow-hll}
\end{center}
\end{figure}

\begin{figure}[H]
\begin{center}
\includegraphics[width=225pt]{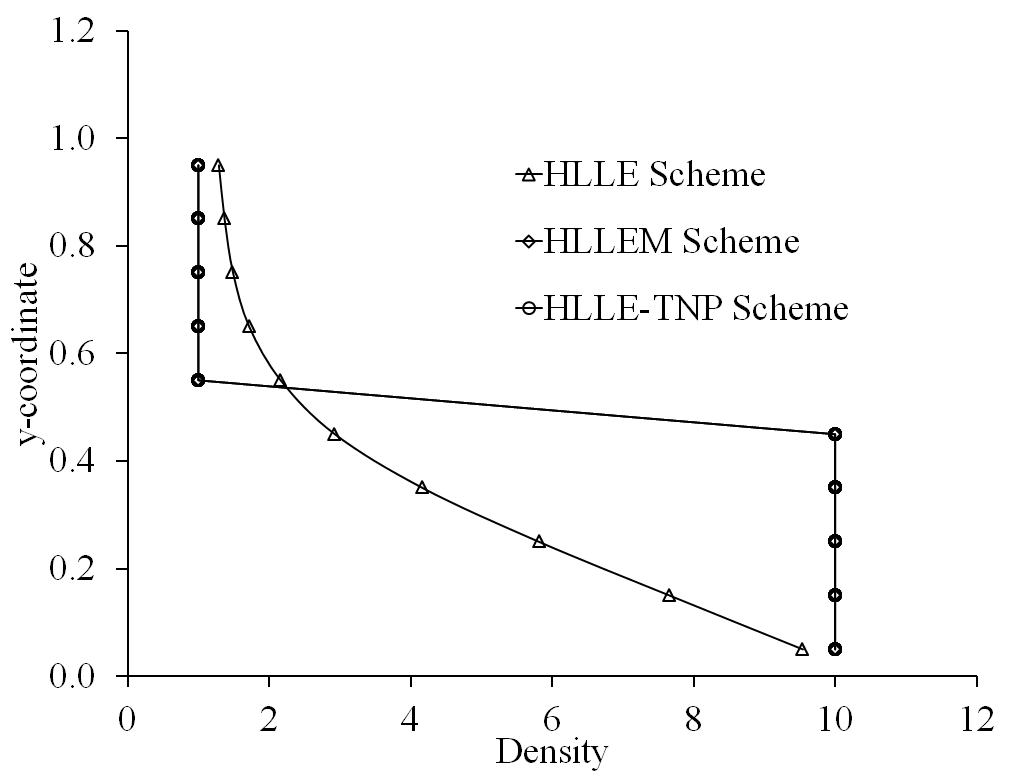}
\label{shear-layer-comparison}
\caption{Density variation along the y-axis at x=0.5 computed by the HLLE and HLLE-TNP schemes for inviscid shear flow.} 
\end{center}
\end{figure}

This test has been proposed by Wada and Liou \cite{wada}  and is used to assess the ability of the numerical scheme to resolve  the shear layers.  The test setup consists of two fluids with different densities sliding at different speeds over each other. The conditions for the top and bottom fluids were chosen as $(\rho, p, M)_{top} = (1.0, 1.0, 2.0)$  and $(\rho, p, M)_{bottom} = (10.0, 1.0, 1.1)$ respectively. A coarse grid consisting of $10\times 10$ cells on a domain of $1.0 \times 1.0$ is used. The CFL number for the simulations were taken to be 1.0, and simulations were run for 1000 iterations. All simulations were computed to first order accuracy. To clearly distinguish between the behaviors of shear  preserving and shear dissipative schemes, the solution for the HLLE scheme is also given for comparison. The density contour plots of the HLLE and the proposed HLLE-TNP scheme is shown in Fig. \ref{shear-flow-hll}. It can be seen from the figure that the proposed HLLE-TNP scheme is able to preserve the shear layer while the  shear layer is dissipated by the HLLE scheme.  The  density plot computed by the HLLE and HLLE-TNP schemes along the y-axis at mid  section (x=0.5) is shown in  Fig. \ref{shear-layer-comparison}. It can be seen from the figure that the HLLE-TNP is able to preserve the shear layer  while the shear layer of the HLLE scheme is diffused.
\subsection{Planar shock problem}
\begin{figure}[H]
	\begin{center}
		\includegraphics[width=250pt]{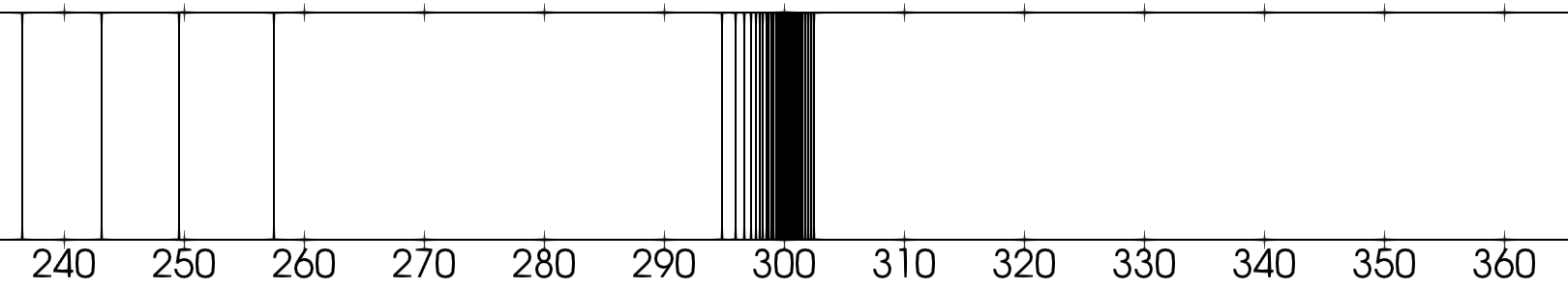} \\ (a) HLLE  Scheme\\
		\includegraphics[width=250pt]{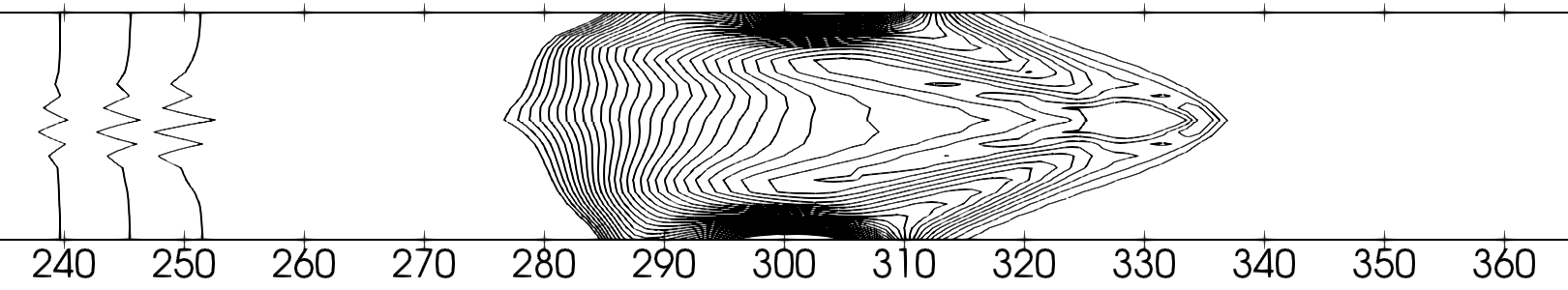} \\ (b) HLLEM Scheme\\
		\includegraphics[width=250pt]{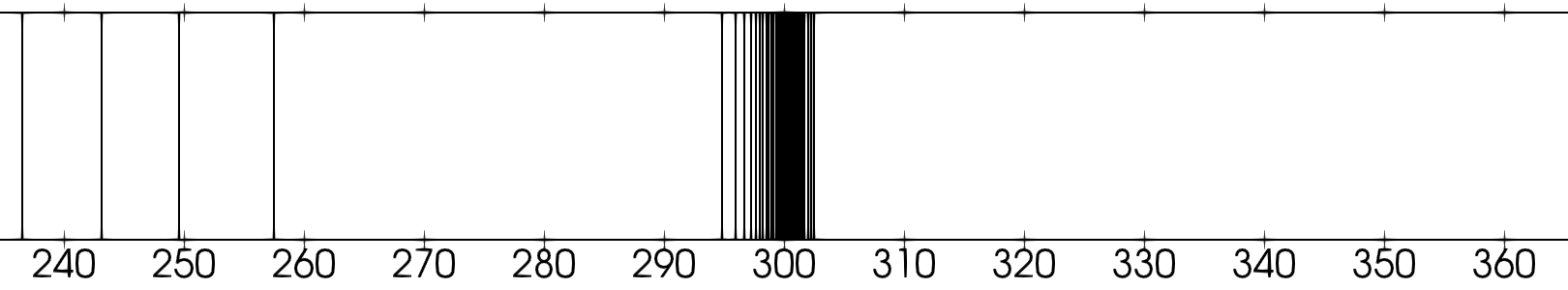} \\ (c) HLLE-TNP  Scheme\\ 
		\caption{Density contours for the $M_{\infty}=6$ Planar Shock problem computed by the HLLE scheme, HLLEM scheme and the HLLE-TNP scheme. The results are shown at t=55 units.}
		\label{planar-shock-hll}
	\end{center}
\end{figure}

The problem consists of a Mach 6.0 shock wave propagating down in a rectangular channel. The domain consists of 800$\times{}$20 cells. The center-line in y direction (11$^{th}$ grid line) is perturbed to promote odd-even decoupling along the length of the shock as
	$y_{i,11}=\left\lbrace\begin{array}{c}\ y_{i,11}+0.001 \hspace{5mm}\text{if  i  is  even}\\ 
	y_{i,11}-0.001\hspace{5mm}\text{if  i  is  odd}\end{array}\right.$. The domain is initialized with  $\rho=1.4,p=1.0, u=0, v=0$. Post shock values are imposed at inlet and zero gradient boundary condition are imposed at exit. Solid wall boundary conditions are imposed at top and bottom. Density contour plots are shown in Fig. \ref{planar-shock-hll} for the HLLE, HLLEM and HLLE-TNP schemes at time t=55 and 30 contour levels ranging from 16 to 7.0 are shown. It can be seen from figure that odd-even decoupling is absent in the proposed HLLE-TNP,  just like the original HLLE scheme.  A severe odd-even decoupling problem is observed in the density contour plot of the  HLLEM scheme.
\subsection{Double Mach reflection problem}
\begin{figure}[H]
	\begin{center}
		\includegraphics[width=250pt]{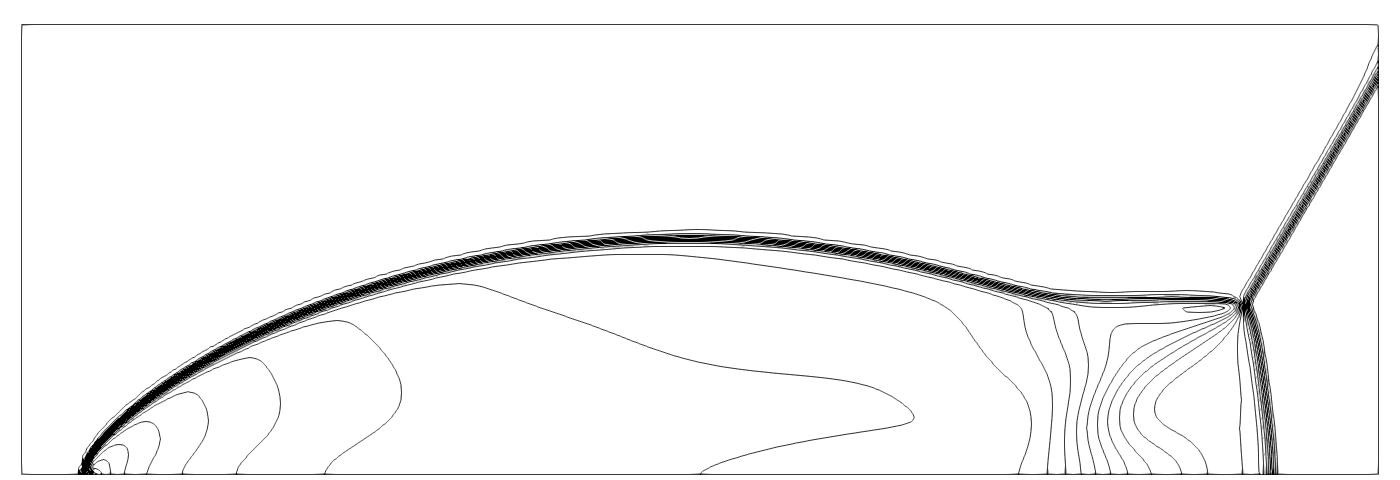} \\ (a) HLLE Scheme\\
		\includegraphics[width=250pt]{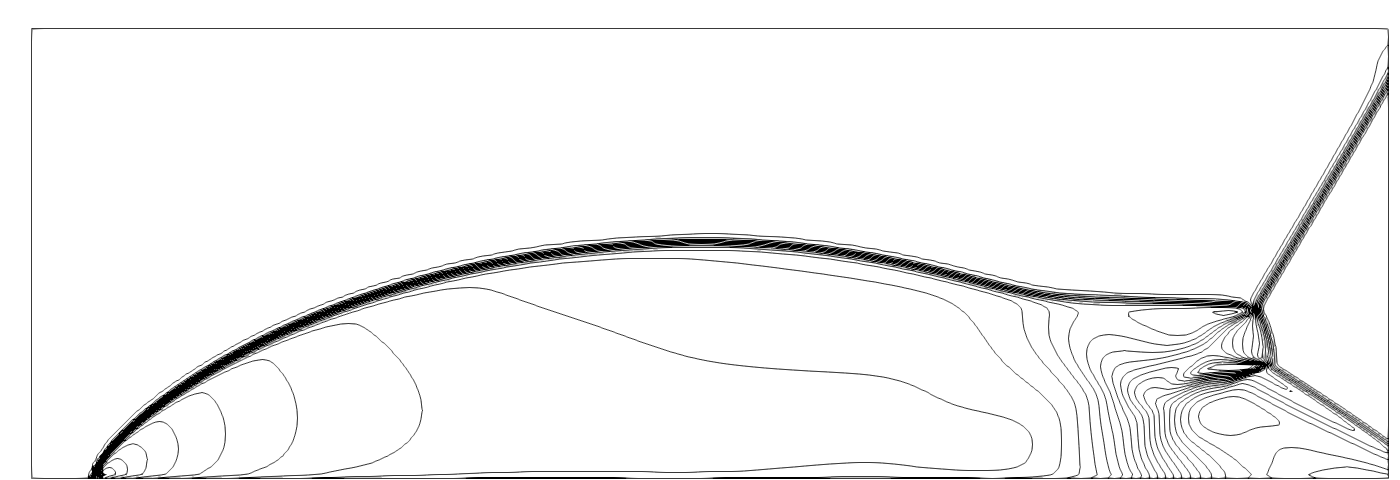} \\ (b) HLLEM  Scheme \\
		\includegraphics[width=250pt]{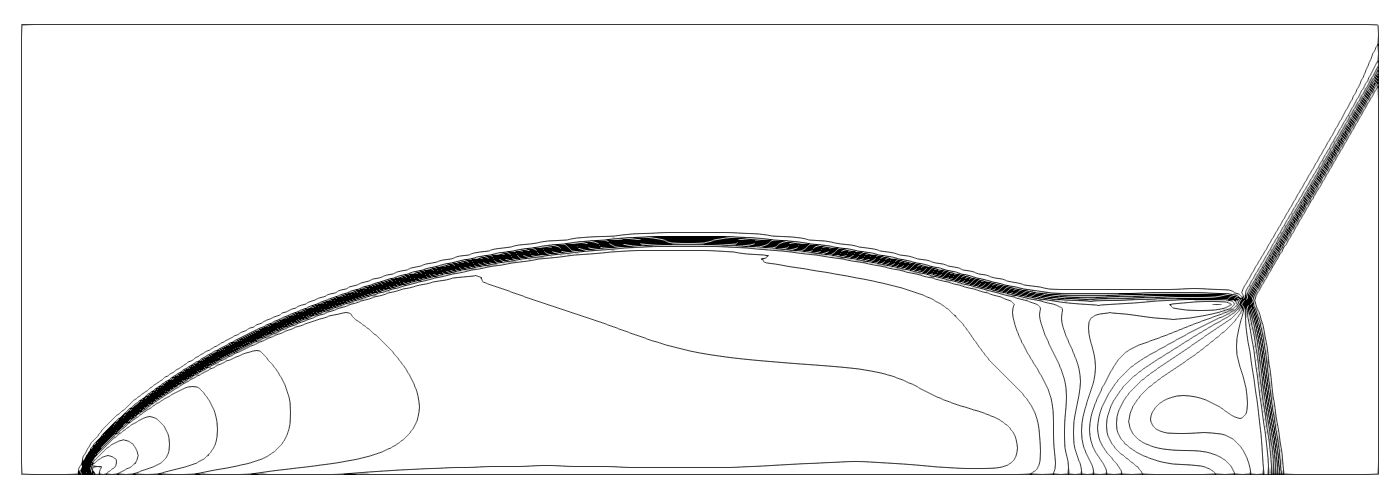} \\ (c) HLLE-TNP Scheme
		\caption{Density contours for the $M_{\infty}=10$ Double Mach Reflection problem computed by the HLLE scheme, HLLEM scheme and the HLLE-TNP scheme. The results are shown at t=0.20}
		\label{dmr-hlle}
	\end{center}
\end{figure}

The domain is four units long and one unit wide. The domain is divided into 480$\times{}$120 cells.  A Mach 10 shock initially making 60 degree angle at x=1/6 with bottom reflective wall is made to propagate through the domain. The domain ahead of shock is initialized to pre-shock value $(\rho=1.4, p=1, u=0, v=0)$ and domain behind shock is assigned post-shock values. The inlet boundary condition is set to post shock value and outlet boundary is set to zero gradient. The top boundary is set to simulate actual shock movement. At the bottom, post-shock boundary condition is set up to x=1/6 and reflective wall boundary condition is set thereafter. Density contours are shown in Fig. \ref{dmr-hlle} for the HLLE schemes at time t=2.00260$\times{}$10 $^{-1}$ units. Thirty (30) contour levels ranging from 2.0 to 21.5 are shown. It can be seen from figure that modified HLLE-TNP scheme is free from kinked Mach stem, just like the original HLLE scheme. 
\subsection{Forward Facing Step Problem}
\begin{figure}[H]
	\begin{center}
		\includegraphics[width=225pt]{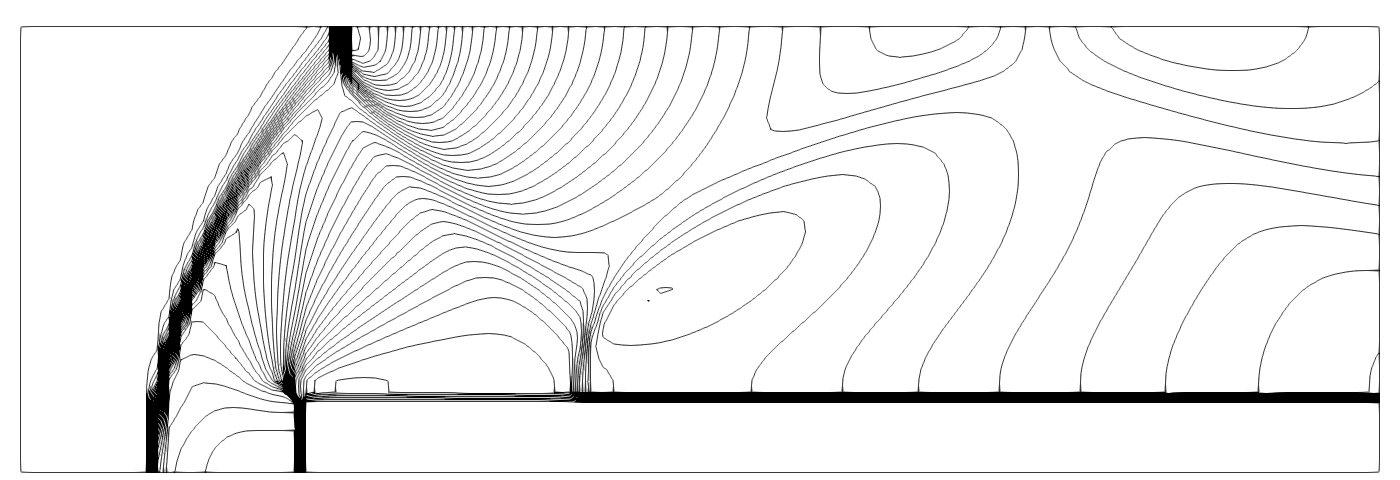} \\(a) HLLE Scheme \\
		\includegraphics[width=225pt]{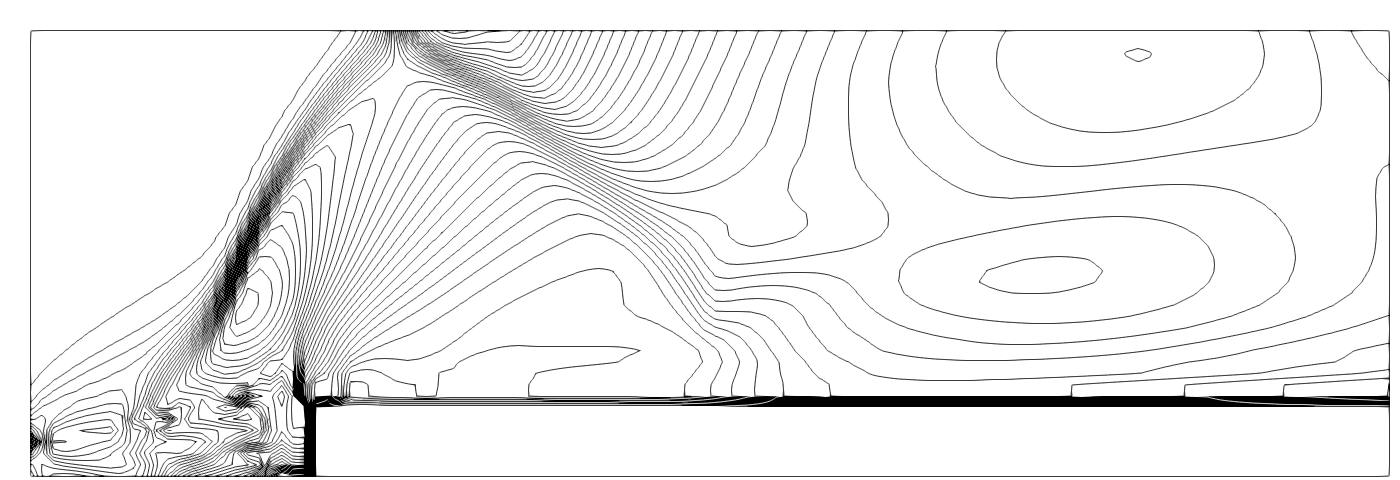} \\(b) HLLEM Scheme \\
		\includegraphics[width=225pt]{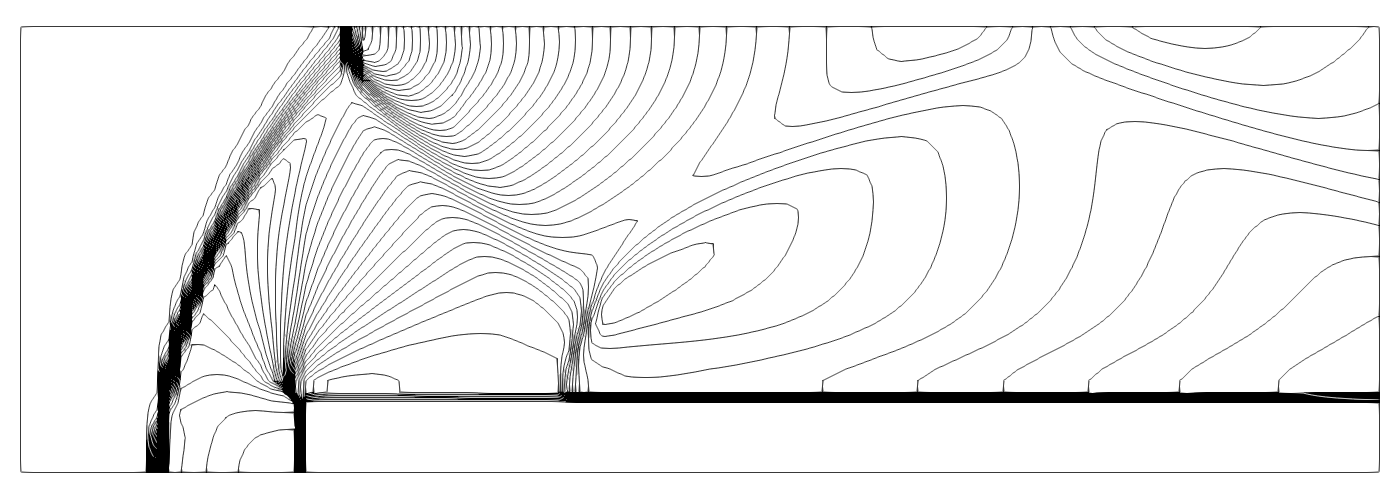} \\ (c) HLLE-TNP Scheme 
		\caption{Density contour for the $M_{\infty}=3$ flow over a Forward Facing Step computed by the HLLE scheme, the HLLEM scheme and the HLLE-TNP scheme. The results are shown at t=4.}
		\label{ffs-hlle}
	\end{center}
\end{figure}

The problem consists of a Mach 3 flow over a forward facing step. The geometry consist of a step located 0.6 units downstream of inlet and 0.2 units high. A mesh of 120$\times{}$40 cells is used for a domain of 3 units long and 1 unit high. The complete domain is initialized with $\rho=1.4, p=1, u=3, v=0$. The inlet boundary is set to free-stream conditions, while outlet boundary has zero gradient. At the top and bottom, reflective wall boundary conditions are set.  Density contour plots for the HLLE schemes are shown in Fig. \ref{ffs-hlle} at time t=4.0 units. A total of 45 contours from 0.2 to 7.0 are shown in the figure. It can be seen from figure that  the proposed HLLE-TNP scheme is able to capture primary and reflected shock without numerical instabilities. 
\subsection{Supersonic flow over a blunt body }
\begin{figure}[H]
	\begin{center}
		\includegraphics[width=90pt]{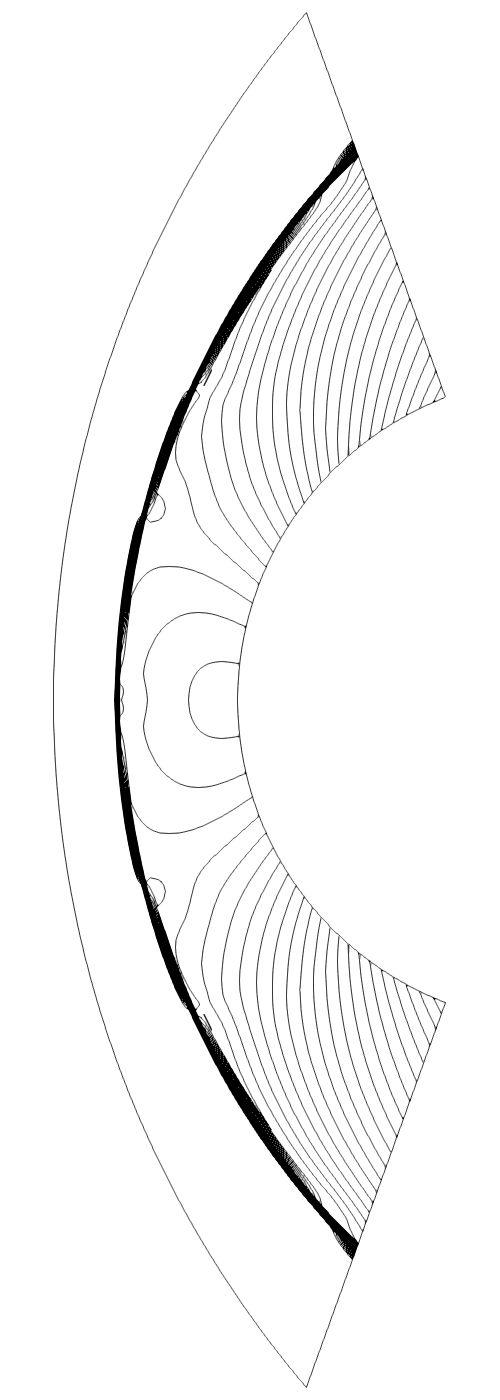} \includegraphics[width=90pt]{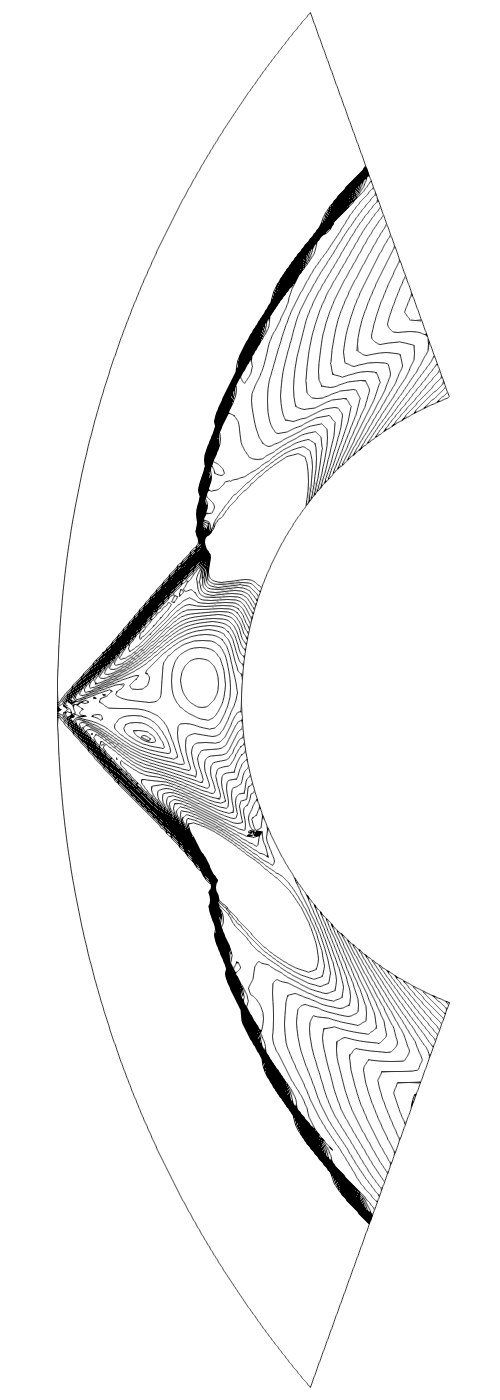} 
		\includegraphics[width=90pt]{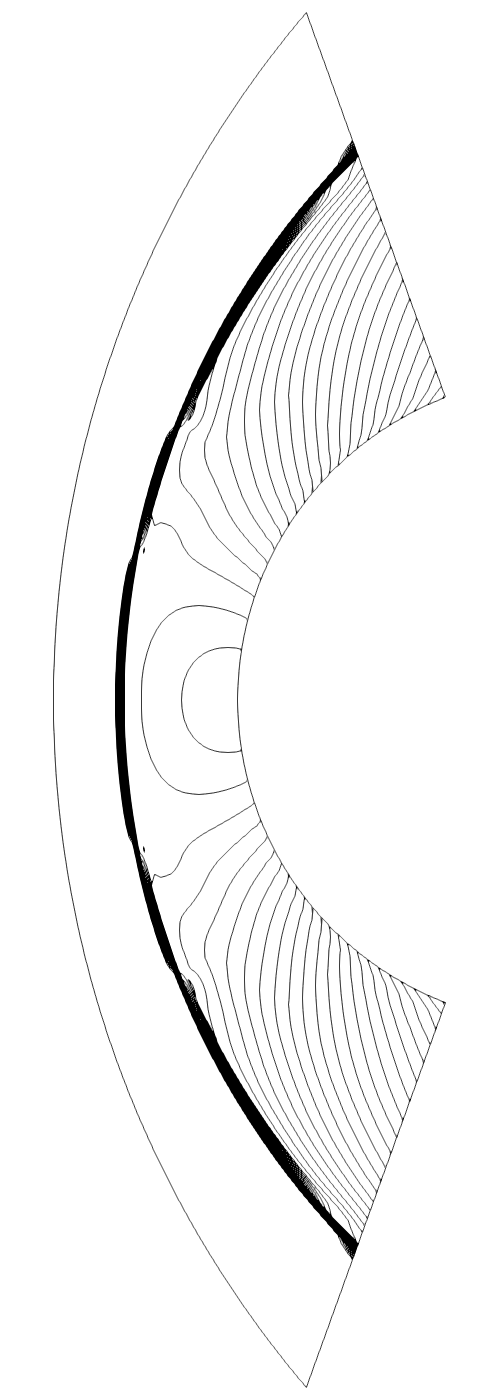} \\
		(a) HLLE Scheme \hspace{7 mm} (b)HLLEM Scheme \hspace{7 mm} (c) HLLE-TNP Scheme 
		\caption{Density Contours for the $M_{\infty}=20$ flow over a blunt body computed by the HLLE scheme, the HLLEM scheme and the HLLE-TNP scheme. The results are shown after 100,000 iterations}
		\label{carbuncle-hlle}
	\end{center}
\end{figure}

\begin{figure}[H]
	\begin{center}
	\includegraphics[width=225pt]{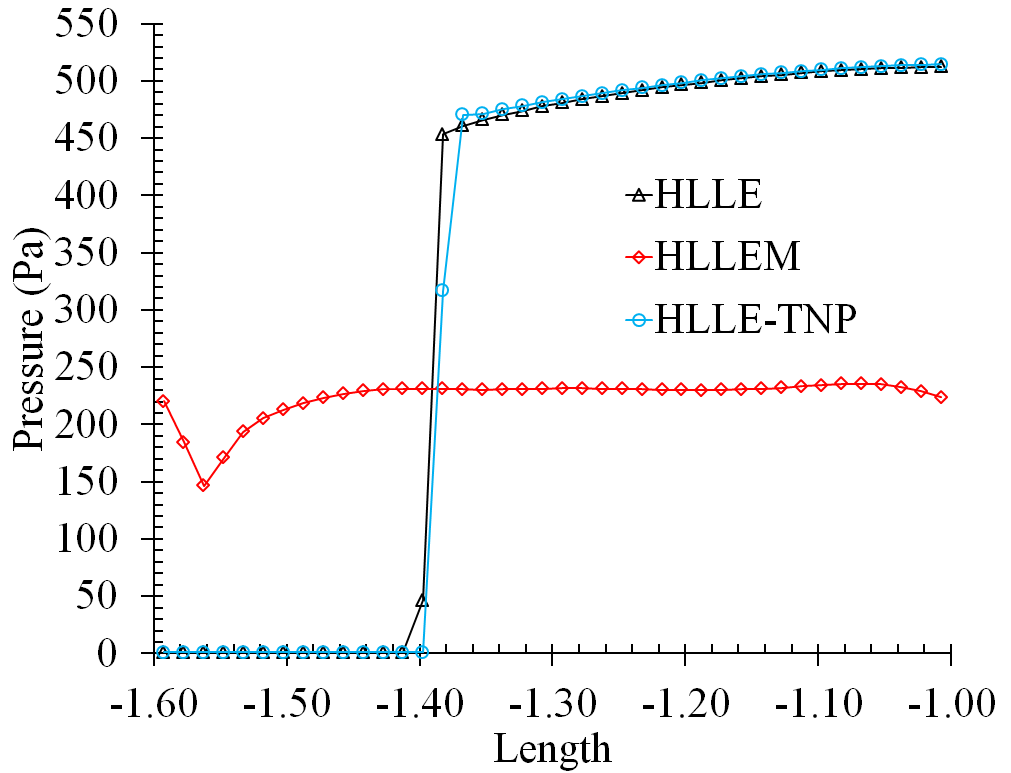} \\ (a) Overall\\ 
	\includegraphics[width=225pt]{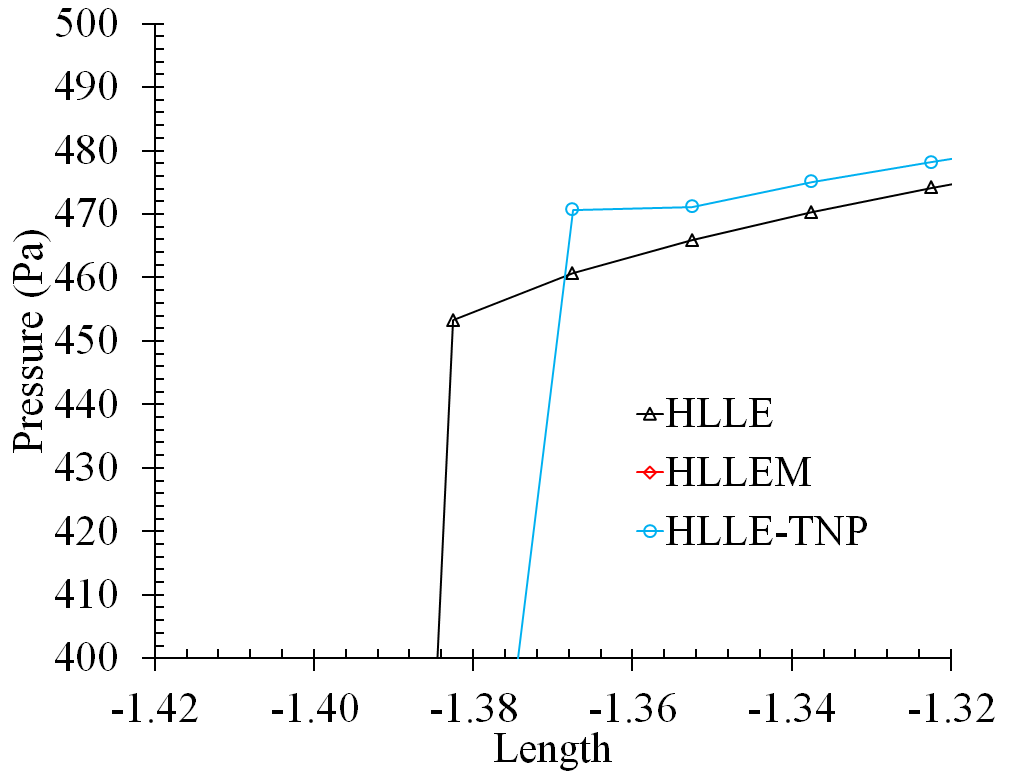}  \includegraphics[width=225pt]{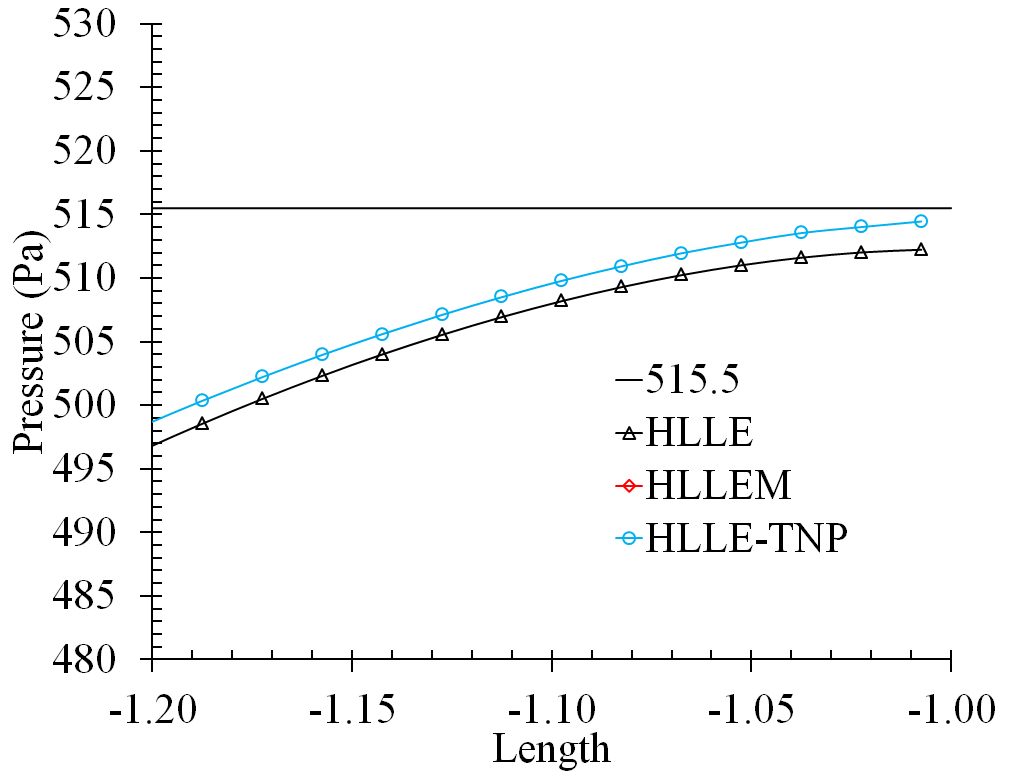}  \\
	(b) Post Shock \hspace{5cm} (c)  Near Stagnation Point
		\caption{Centerline static pressure Plot for the $M_{\infty}=20$ flow over a blunt body computed by the HLLE scheme, the HLLEM scheme and the HLLE-TNP scheme. The results are shown after 100,000 iterations}
		\label{centerline-carbuncle-hlle}
	\end{center}
\end{figure}

\begin{figure}[H]
	\begin{center}
		\includegraphics[width=225pt]{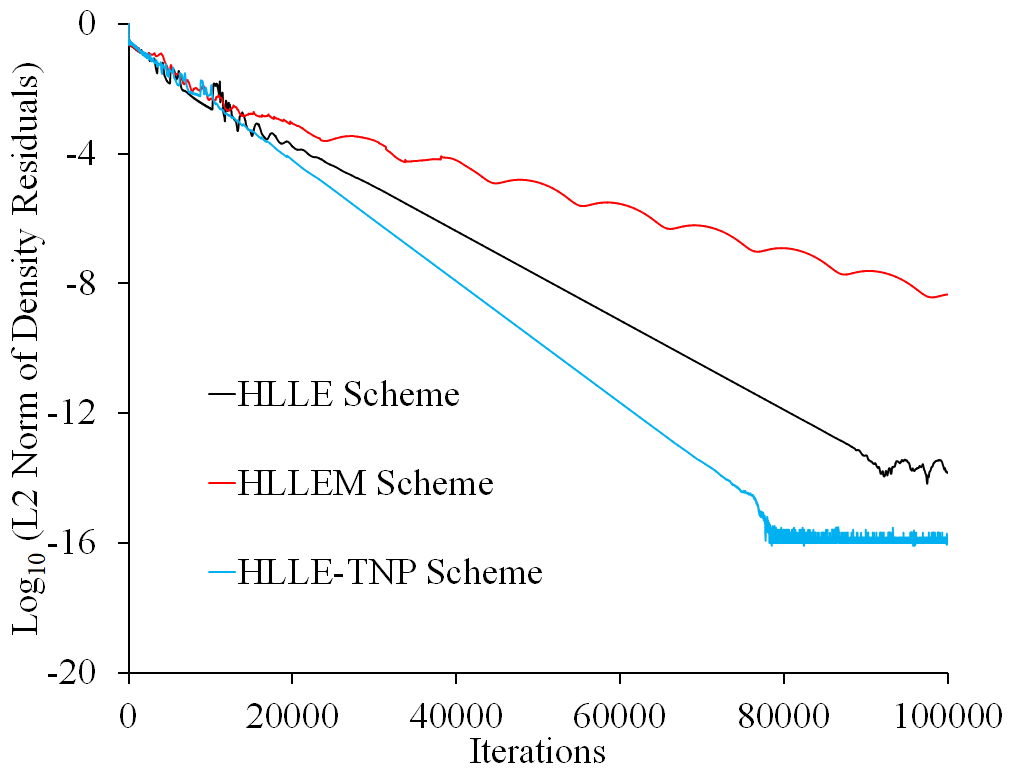} 
		\caption{History of the L-2 norm of density residual for the $M_{\infty}=20$ flow over a blunt body computed by the HLLE scheme, the HLLEM scheme and the HLLE-TNP scheme}
		\label{convergence-carbuncle-hlle}
	\end{center}
\end{figure}

Supersonic flow over a blunt body is a typical problem to assess the performance of a scheme with respect to the carbuncle instability. Free-stream Mach number of 20 is considered in the computation. The grid for the blunt body  had a size of 40 $\times{}$ 320 cells. The domain is initialized with values of $\rho=1.4, p=1, u=20, v=0$.  Inlet boundary condition is set to free-stream value. Solid wall boundary condition is applied on the blunt body. The contour plot for density for the HLLE, HLLEM and HLLE-TNP schemes are shown in Fig. \ref{carbuncle-hlle}, and a total of 27 density contours from 2.0 to 8.7 are drawn. It can be seen from the figure that the original HLLE and the proposed HLLE-TNP schemes are free from the carbuncle phenomenon, while  the HLLEM scheme exhibit a severe carbuncle phenomenon. The static pressure plot along the center-line is shown in Fig. \ref{centerline-carbuncle-hlle}. It can be seen from the figure that the classical HLLE and the HLLE-TNP scheme produce monotone solution  without any overshoot or undershoot. 
The post-shock and stagnation static pressure  for the various HLLE-type schemes is shown in Table \ref{hlle-pressure}. It can be seen from the table that the static pressure at the stagnation point of the proposed HLLE-TNP  scheme is closer to the analytical value than the classical  HLLE scheme. Therefore, it is felt that the low Mach corrections to the approximate Riemann solvers can  contribute to more accurate solutions in the post-shock region of high speed flows  where the Mach number is subsonic. The convergence history of the  HLLE, HLLEM  and the proposed HLLE-TNP  scheme are shown in Fig. \ref{convergence-carbuncle-hlle}. It can be seen from the figure that  the proposed HLLE-TNP scheme converge to machine accuracy within 100,000 iterations, while the residual of the classical HLLE and HLLEM  schemes converges to about $10^{-14}$ and $ 10^{-8}$ respectively.
\begin{table}[H]
\begin{doublespace}
\begin{center}
\caption {Static Pressure Comparison of HLLE-type schemes for flow past a blunt body}
\label{hlle-pressure}
\begin{tabular}{|l|l|l|l|l|}
\hline Serial No & Scheme & Post-Shock Pressure (Pa)& Stagnation Pressure (Pa) \\ 
\hline  1 & Analytical & 466.5 & 515.5\\ 
\hline  2 & HLLE & 460.67 & 512.23\\   
\hline  3 & HLLE-TNP & 470.63 & 514.45\\ 
\hline
\end{tabular}
\end{center}
\end{doublespace}
\end{table}
\subsection{Diffraction of a moving normal shock over a $90^o$ corner}
\begin{figure} [H]
	\begin{center}
		\includegraphics[width=200pt]{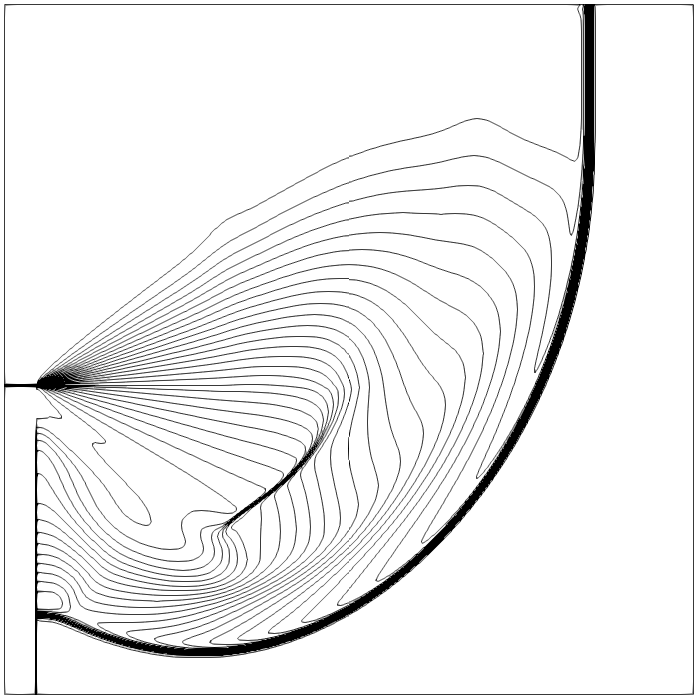} 
		\includegraphics[width=200pt]{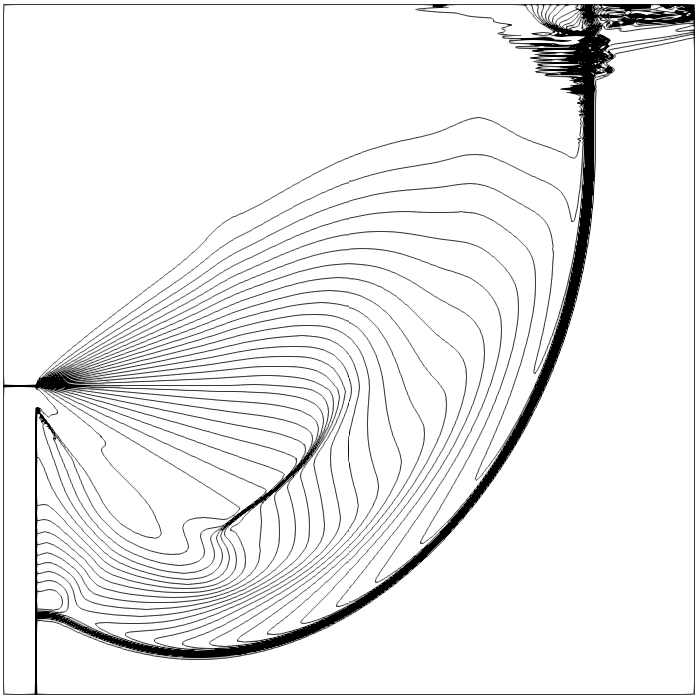} 
		\\ (a) HLLE Scheme \hspace{3cm} (b) HLLEM Scheme \\ 
		\includegraphics[width=200pt]{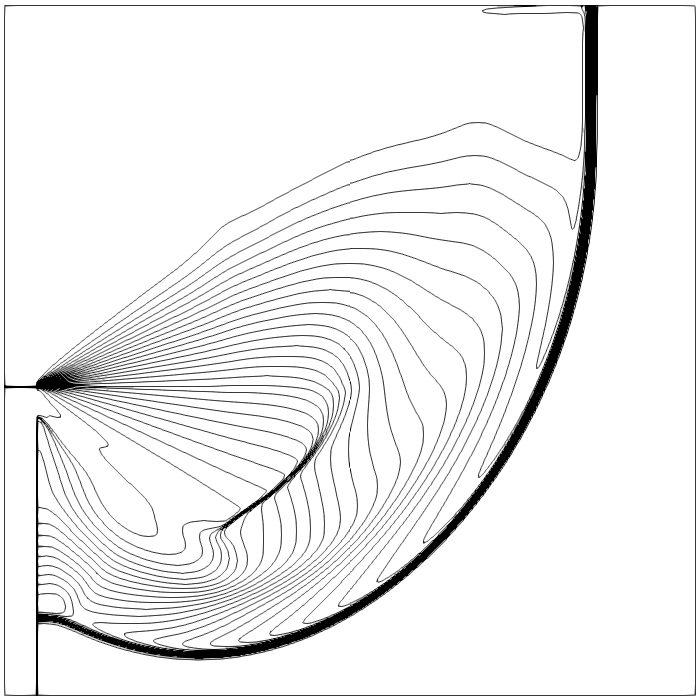}
		\\ (c) HLLE-TNP Scheme  \\ 
		\caption{Density Contours for  the $M_{\infty}=5.09$ normal shock diffraction around a $90^o$ corner computed by the HLLE scheme, the HLLEM scheme and the HLLE-TNP scheme. The results are shown at t=0.1561.}
		\label{scr-hlle}
	\end{center}
\end{figure}

The problem consists of sudden expansion of Mach 5.09 normal shock around a 90 degree corner. The domain is a square of one unit and is divided into 400$\times{}$ 400 cells. The corner is located at x=0.05 and y=0.45. The initial normal shock is located at x=0.05. The domain to the right of shock is assigned initially with pre-shock conditions of $\rho{}$=1.4, p=1.0, u=0.0, v=0.0. The domain to left of shock is assigned post shock conditions. The inlet boundary is supersonic, outlet boundary has zero gradient and bottom boundary behind the corner uses extrapolated values. Reflective wall boundary conditions are imposed on the corner. A CFL value of 0.8 is used and plot is generated for time t=0.1561 units.  A total of 30 density contours ranging from 0 to 7.1 is shown in  Fig. \ref{scr-hlle} for the  HLLE schemes. The figure shows that the classical HLLE scheme and the proposed HLLE-TNP schemes are free from the  numerical shock instability problem, while shock instability is observed in the HLLEM scheme.
\subsection{Inviscid Flow past a circular cylinder at low Mach number }
\begin{figure} [H]
	\begin{center}
		\includegraphics[width=150pt]{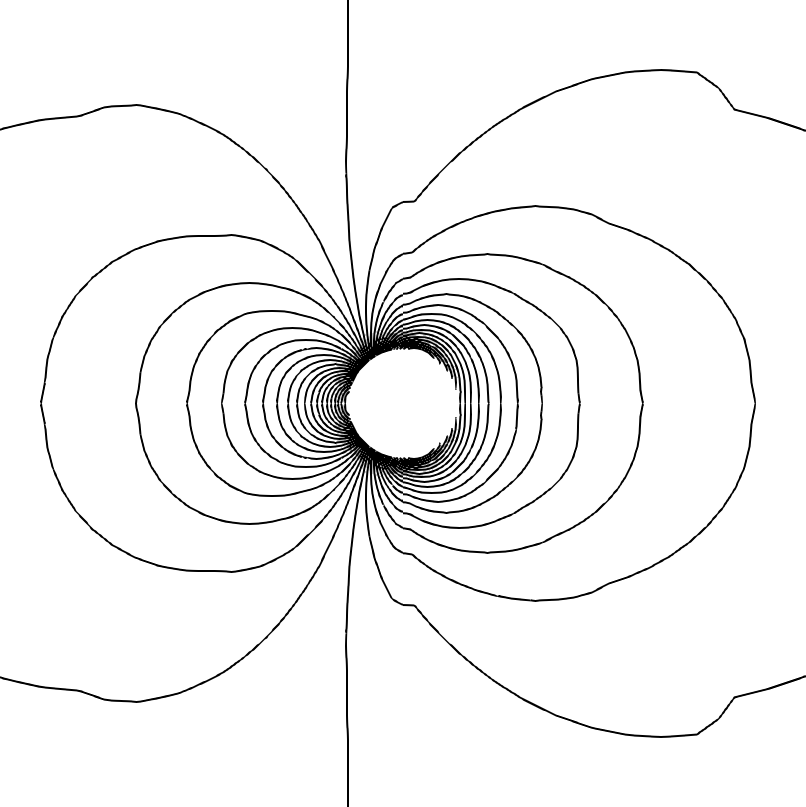}  
		\includegraphics[width=150pt]{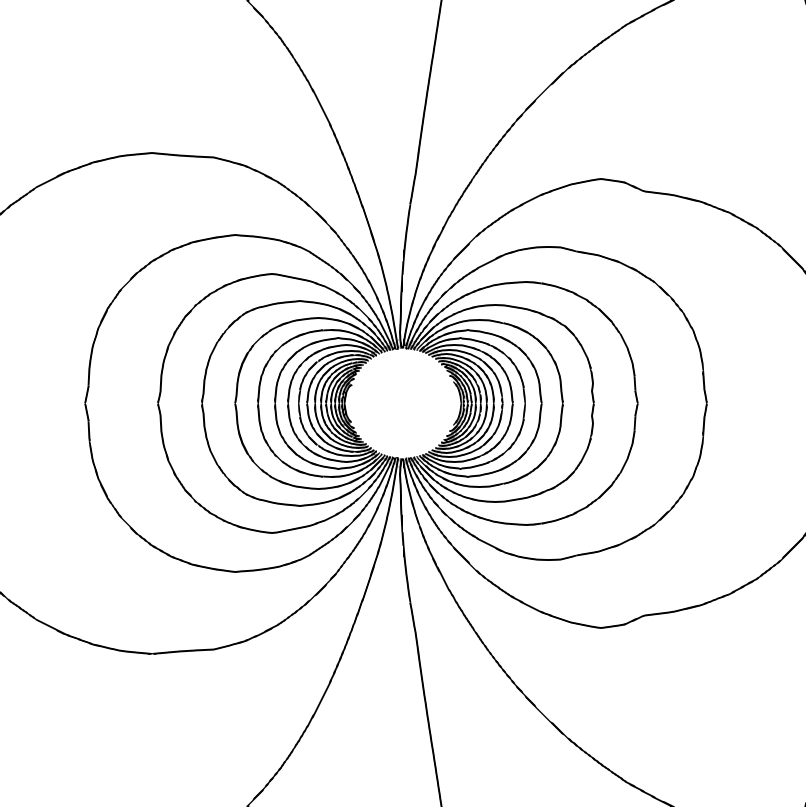} 			
		\includegraphics[width=150pt]{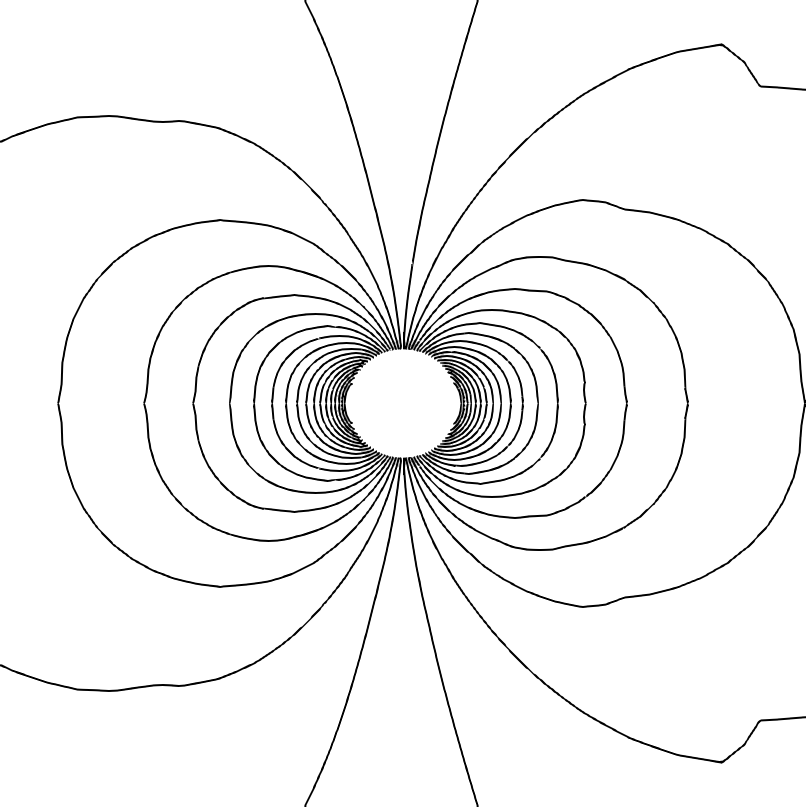} \\
		(a) Mach 0.100 \hspace{3cm}  (b) Mach 0.010 \hspace{3cm}	(c) Mach 0.001 
		\caption{Static Pressure Contour plot for  flow past a circular cylinder computed by the HLLE Scheme}
		\label{cylinder-hlle}
	\end{center}
\end{figure}

\begin{figure} [H]
	\begin{center}
		\includegraphics[width=150pt]{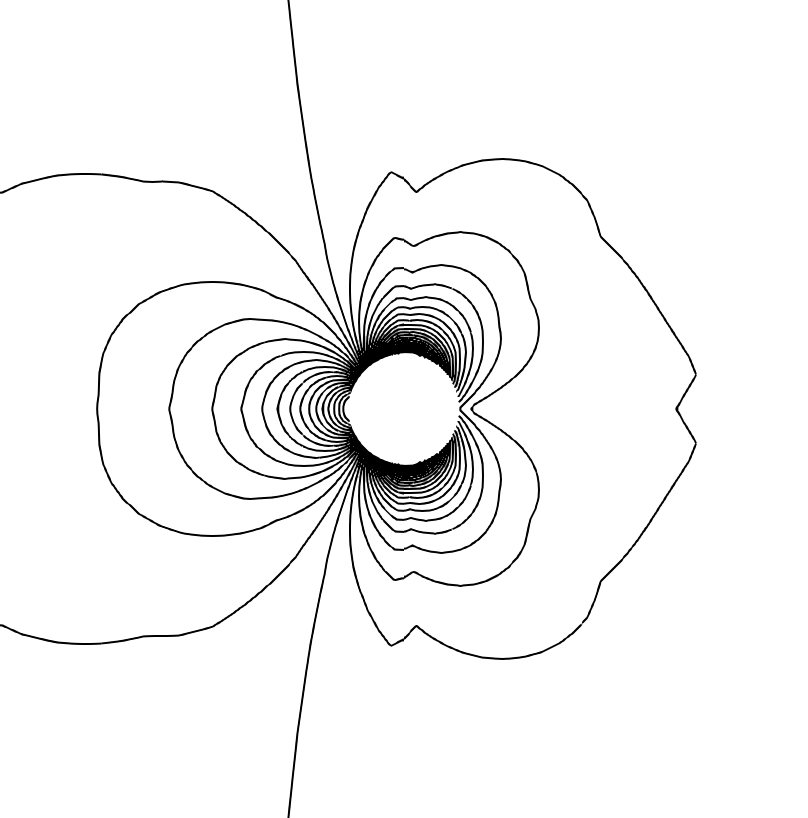}  
		\includegraphics[width=150pt]{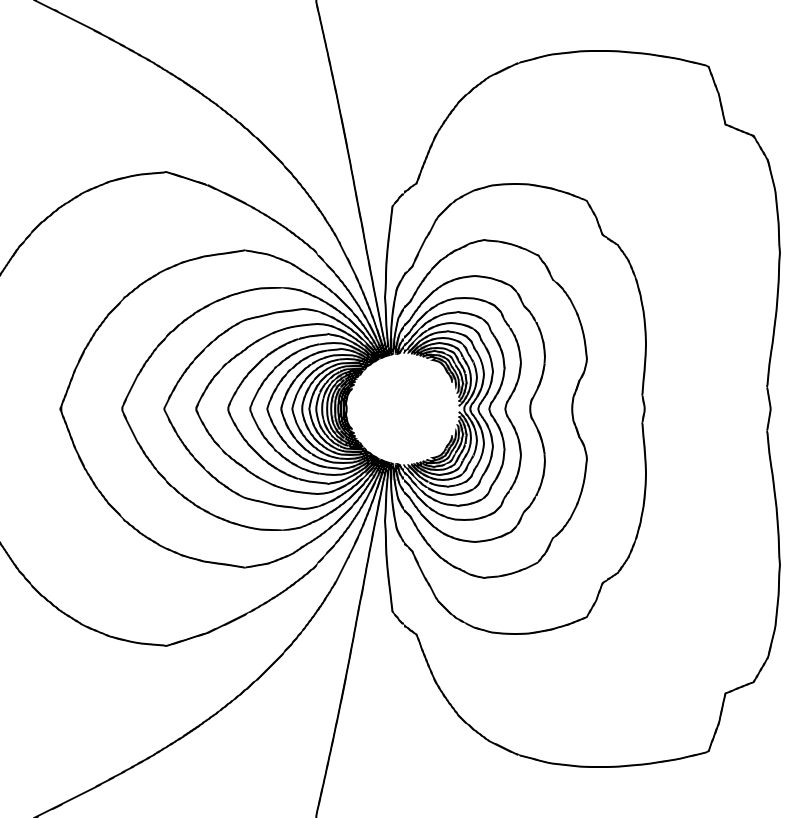} 			
		\includegraphics[width=150pt]{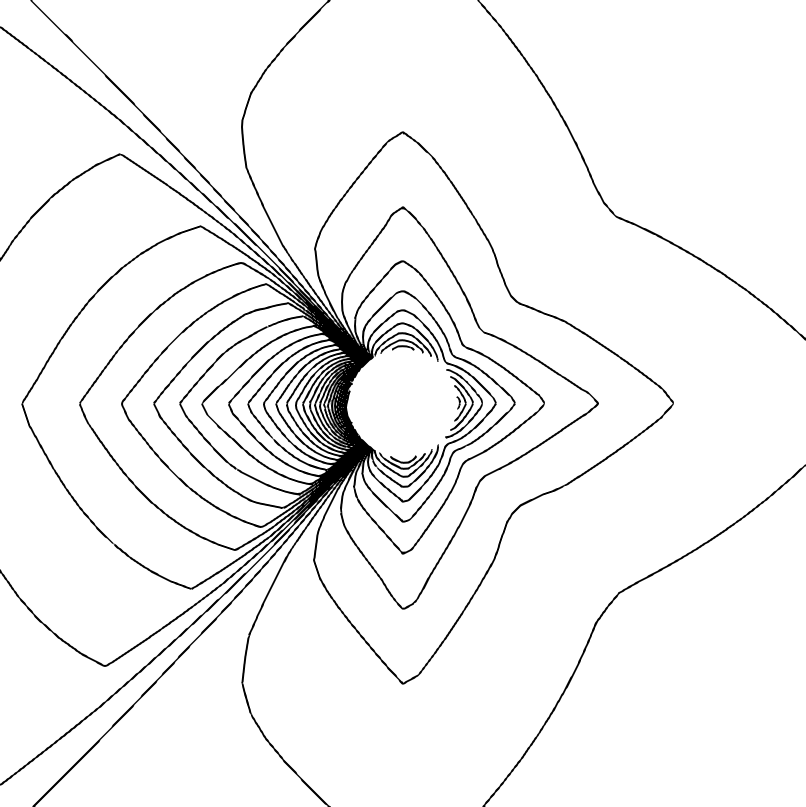} \\
		(a) Mach 0.100 \hspace{3cm}  (b) Mach 0.010 \hspace{3cm}	(c) Mach 0.001 
		\caption{Static Pressure Contour plot for  flow past a circular cylinder computed by the HLLEM Scheme}
		\label{cylinder-hllem}
	\end{center}
\end{figure}

\begin{figure}[H]
	\begin{center}	
		\includegraphics[width=150pt]{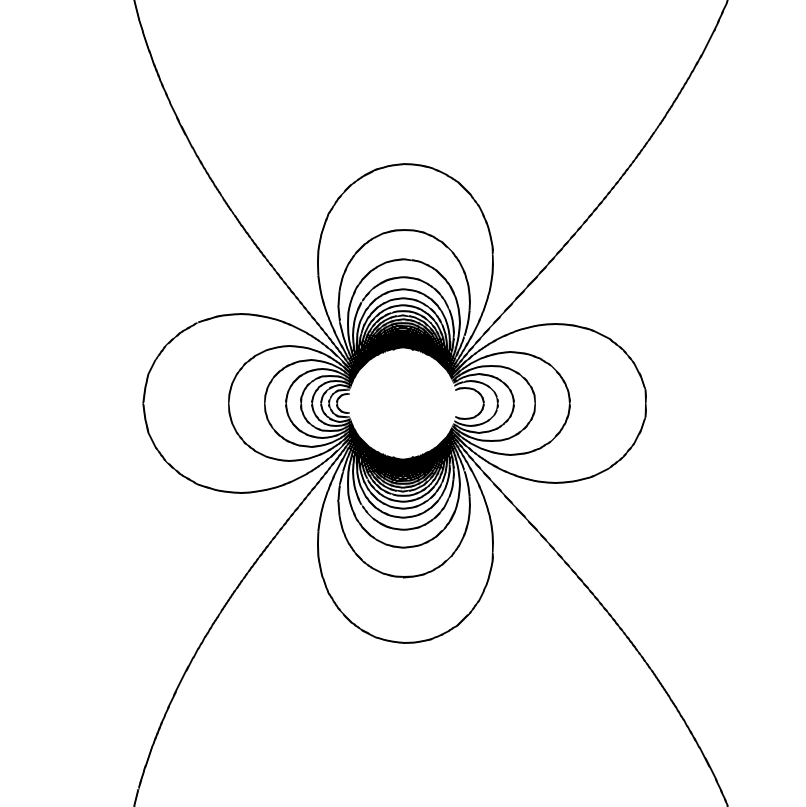}  
		\includegraphics[width=150pt]{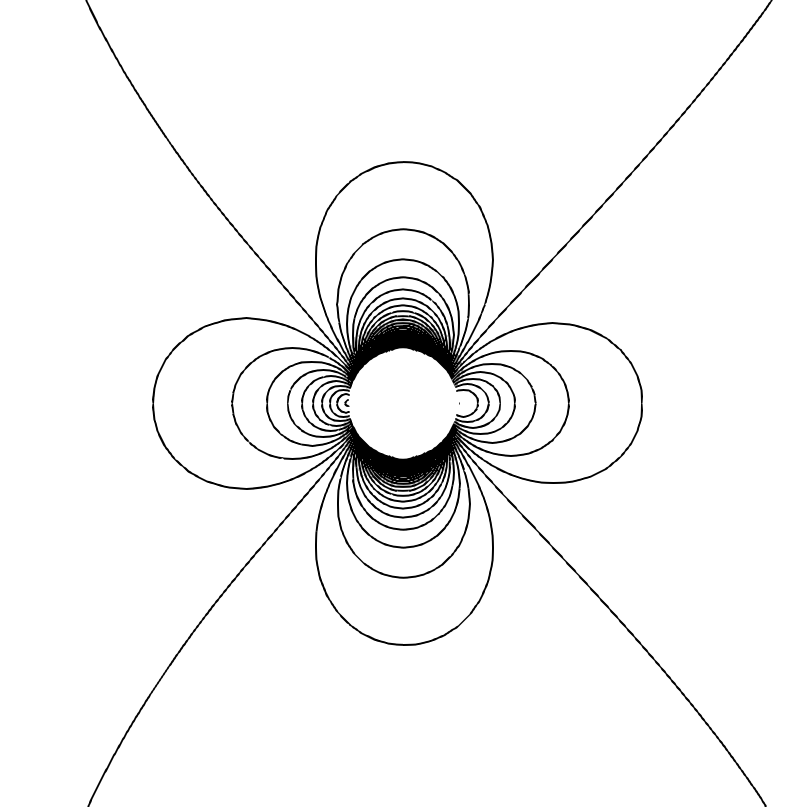} 
		\includegraphics[width=150pt]{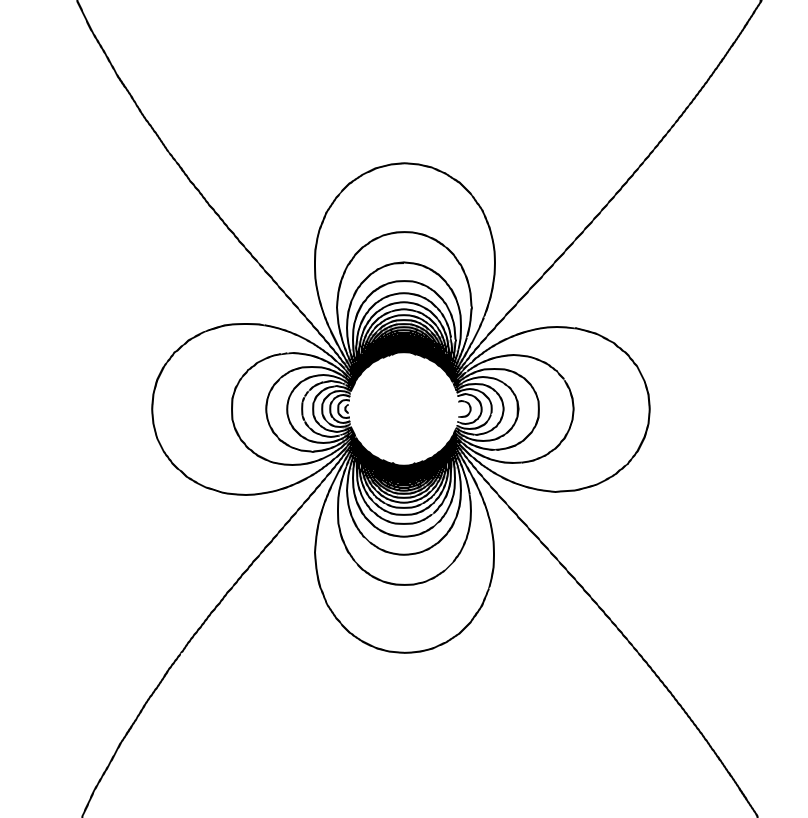} \\
		(a) Mach 0.100 \hspace{3cm}	(b) Mach 0.010 \hspace{3cm}	(c) Mach 0.001 
		\caption{Static Pressure Contour plot for  flow past a circular cylinder computed by  the proposed HLLE-TNP Scheme}
		\label{cylinder-HLLE-TNP}
	\end{center}
\end{figure}

\begin{figure}[H]
	\begin{center}	
		\includegraphics[width=225pt]{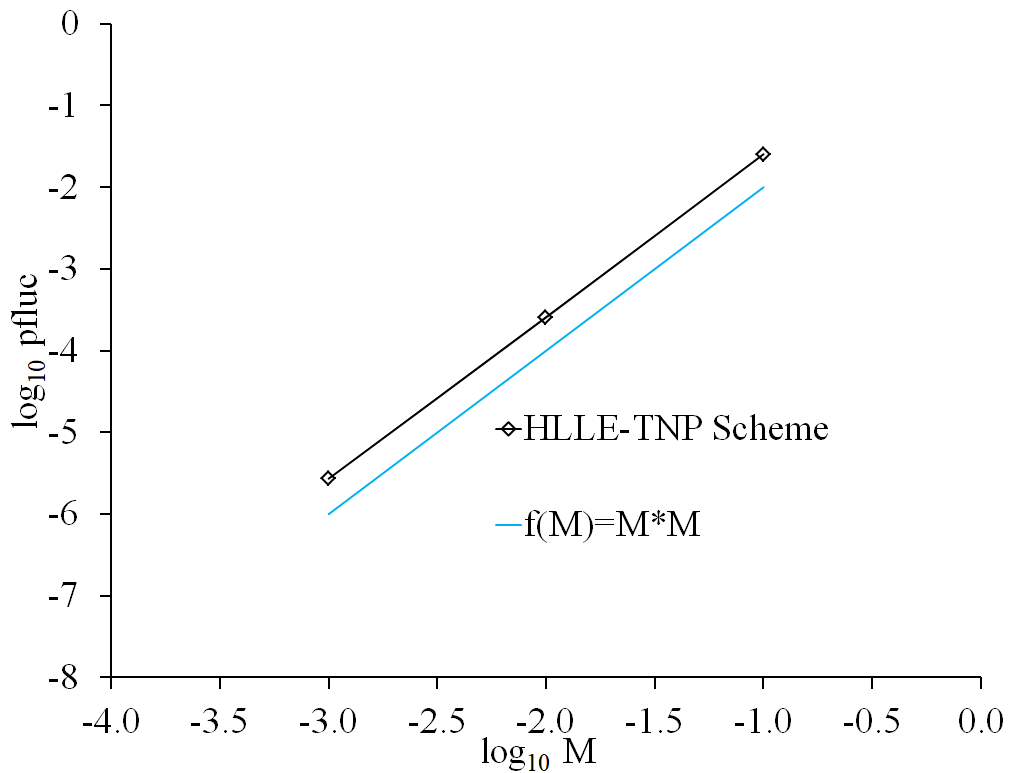} 
		\caption{Maximum pressure fluctuation $p_{fluc}=(p_{max}-p_{min})/p_{max}$ against inflow Mach number for the flow around a cylinder obtained with the proposed HLLE-TNP scheme}
		\label{cylinder-pfluc-HLLE-TNP}
	\end{center}
\end{figure}

The circular cylinder at low Mach number is a important test case to ascertain the ability of the scheme to resolve the low Mach flow features. First order, inviscid calculations are carried out on circular cylinder using the  classical HLLE scheme and the proposed HLLE-TNP scheme. The computations are carried out with a CFL number of 0.80. Structured Grids of size 49$\times$ 37, 97$\times$  37, 97$\times$ 73 are used and results are shown for grid size of $97\times73$. The static pressure plots of  the classical HLLE scheme for Mach numbers 0.10, 0.01 and 0.001 are shown in Fig. \ref{cylinder-hlle}. Due to increase in numerical viscosity with reduction in Mach number in HLLE scheme, the pressure contours resemble creeping Stokes flow. The static pressure plot also show highest numerical viscosity for Mach 0.001 case, indicating that numerical viscosity increases with decrease is Mach number for HLLE scheme. 
The static pressure plots of  the HLLEM scheme for Mach numbers 0.10, 0.01 and 0.001 are shown in Fig. \ref{cylinder-hllem}. It can be seen from the figure that the  static pressure plots of the HLLEM scheme do not resemble potential flow, but are somwhat intermediate between Stokes flow and potential flow. The marginally better  results of the HLLEM scheme  in comprison to the HLLE scheme can be attributed to the fact that the HLLEM scheme contain significant velocity diffusion terms only in the normal momentum flux, while the HLLE scheme contain significant velocity diffusion terms in both the normal and transverse momentum flux.  
The static pressure plot of the proposed HLLE-TNP scheme for Mach 0.1, 0.01 and 0.001 are shown in Fig. \ref{cylinder-HLLE-TNP}. Static pressure plot of the modified scheme resemble potential flow even at Mach 0.001. The improved performance of the proposed HLLE-TNP scheme can be attributed to the reduction in numerical dissipation through  the velocity reconstruction method. The maximum pressure fluctuation  in the flow field has been  defined as \cite{guillard1, rieper1} 
\begin{equation} 
p_{fluc}=\dfrac{p_{max}-p_{min}}{p_{max}} 
\end{equation} 
where $p_{max}$ and $p_{min}$ are the maximum and minimum pressure in the flow field.  For incompressible, potential flow past a circular cylinder, the maximum pressure, which is obtained at the stagnation point, is given by $p_{max}=\dfrac{\gamma}{2}M_{\infty}^2$ while the minimum pressure is given by $p_{min}=-\dfrac{3\gamma}{2}M_{\infty}^2$ \cite{rieper1}. Therefore, the maximum pressure fluctuation $p_{fluc}$ in the flow field for incompressible potential flow is given by $p_{fluc}=2\gamma{}M_{\infty}^2$. The maximum pressure fluctuations computed by the proposed HLLE-TNP scheme for different inflow Mach numbers is shown in Table \ref{cylinder-pfluc-HLLE-TNP-table}. It can be seen from the table that the maximum pressure fluctuations computed by the proposed HLLE-TNP scheme is very close to the pressure fluctuations in the incompressible, potential flow. The maximum pressure fluctuation obtained with the proposed HLLE-TNP scheme for different inflow Mach numbers is plotted in Fig. \ref{cylinder-pfluc-HLLE-TNP}. It can be seen from the figure that the pressure fluctuation is of the order $\mathcal{O} (M^2)$ which is consistent with the pressure fluctuation in the continuous Euler equations.
\begin{table}[H]
\begin{doublespace}
\begin{center}
\caption {Pressure fluctuations for flow around a circular computed by the proposed HLLE-TNP scheme.}
\label{cylinder-pfluc-HLLE-TNP-table}
\begin{tabular}{|c|c|c|c|}
\hline Serial No & $M_{\infty}$ & $p_{fluc}$ & $p_{fluc}$ (Potential Flow) \\ 
\hline  1 & $10^{-1}$ & $2.54\times{}10^{-2}$ & $2.8\times{}10^{-2}$\\ 
\hline  2 & $10^{-2}$ & $2.55\times{}10^{-4}$ & $2.8\times{}10^{-4}$\\   
\hline  3 & $10^{-3}$ & $2.71\times{}10^{-6}$ & $2.8\times{}10^{-6}$\\ 
\hline
\end{tabular}
\end{center}
\end{doublespace}
\end{table}
\subsection{Flat Plate Boundary Layer}
\begin{figure}[H]
	\begin{center}	
		\includegraphics[width=225pt]{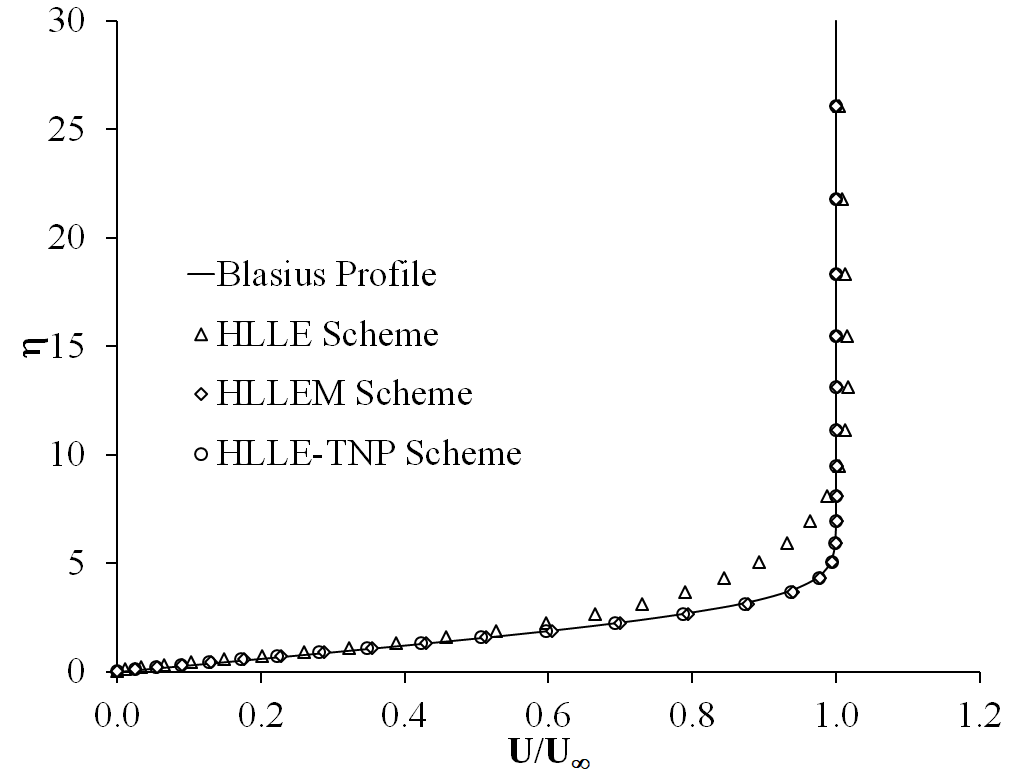}\\
		\caption{Boundary Layer Profile for $M_{\infty}=0.20$ laminar flow over a flat plate using the HLLE, HLLEM and HLLE-TNP schemes.}
		\label{bl-profile-hlle}
	\end{center}
\end{figure}

\begin{figure}[H]
	\begin{center}	
		\includegraphics[width=225pt]{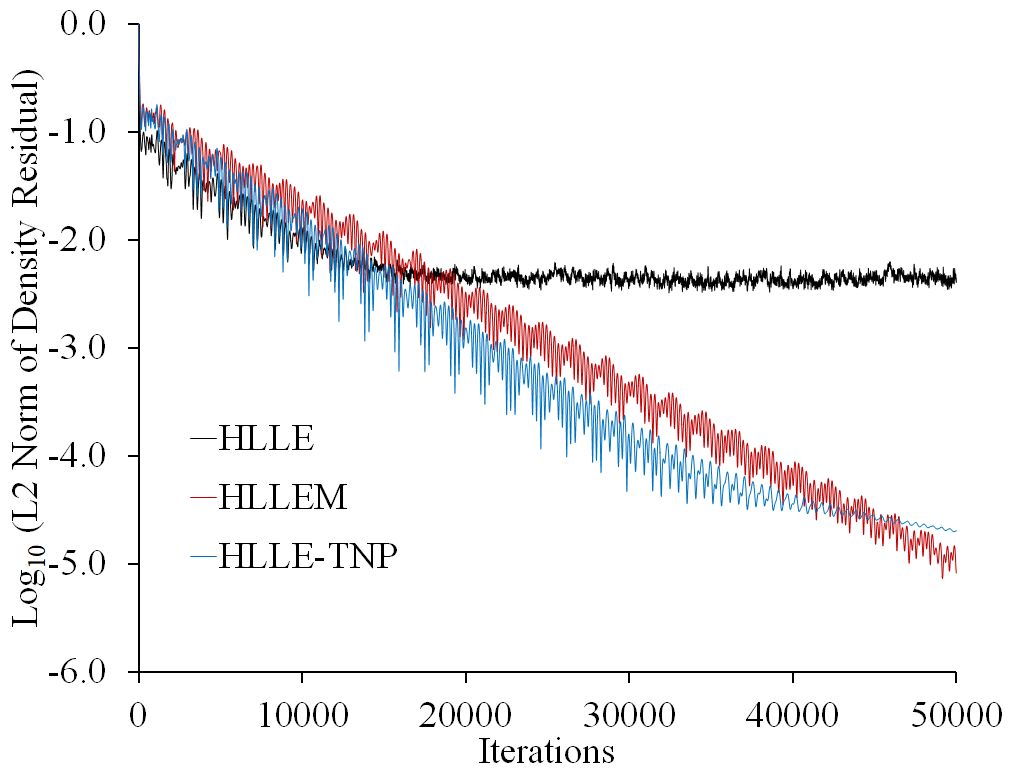}\\
		\caption{History of L2 norm of  density residual for $M_{\infty}=0.20$ laminar flow over a flat plate computed by the  HLLE, HLLEM and HLLE-TNP schemes.}
		\label{bl-convergence}
	\end{center}
\end{figure}

Viscous Computations are carried out with the proposed HLLE schemes  over a flat plate in order to demonstrate the boundary  layer resolving capability of these schemes. The computations are carried out for a freestream Mach number of 0.20   and Reynolds number of about 25,000. The third order spatially accurate computations are carried out with MUSCL approach \cite{leer} without any limiters.  The second order SSPRK method proposed by Gottleib and Shu \cite{gott}  is used for time integration.  The boundary layer profiles are shown after 50,000 iterations with a CFL number of 0.50 on  a grid size of $81 \times 33$. The comparison of the computed boundary layer profiles with the analytical  profile of Blasius is shown in Fig. \ref{bl-profile-hlle}. It can be seen from the figure that   the boundary layer profile  of the  proposed HLLE-TNP and HLLEM schemes are almost identical to the analytical boundary layer profile  of Blasius. The boundary layer profile of the original HLLE scheme  does not match with the analytical  profile of Blasius. The convergence history of density for the different schemes is shown in Fig. \ref{bl-convergence}. It can be seen from the figure that the  residual of density for HLLEM  and HLLE-TNP schemes converge to about $10^{-5}$ while the density residual of the original HLLE scheme stalls at about $10^{-3}$.

\section{Conclusion}
An all Mach number HLLE-type scheme is proposed here which is capable of resolving low Mach flow features and the linearly degenerate waves while preserving shock stability. In the proposed scheme, a  simple velocity reconstruction method based on the face normal Mach number is introduced in the HLLE scheme for resolving the shear layers and the flow features at low Mach numbers. The face normal Mach number function is scaled with a pressure function for preserving shock stability. The resolution of the contact discontinuity is achieved in a manner similar to the HLLEM scheme with  a modified anti-diffusion coefficient for preserving shock stability.
 The ability of the proposed scheme to resolve the isolated contact and shear waves exactly is proven analytically. The capability of the proposed HLLE-TNP scheme for resolving low Mach number flow features is demonstrated analytically through asymptotic analysis. The stability of the proposed  scheme in presence of strong shock is demonstrated  through linear  perturbation and matrix stability analyses. A set of numerical test cases are solved to demonstrate the capabilities of the proposed scheme. The proposed HLLE-TNP scheme is found to be free from the numerical shock instability problems like odd-even decoupling, kinked Mach step and carbuncle phenomenon. For the low Mach number circular cylinder case, the proposed HLLE-TNP scheme is  found to produce results resembling potential flow even at a very low Mach number of 0.001. The proposed HLLE-TNP  scheme is able to exactly resolve inviscid shear layers and accurately resolve flat plate boundary layers.  It is thus demonstrated that the proposed  HLLE-TNP scheme  is capable of resolving the linearly degenerate waves without facing any numerical instability at high Mach number  and is capable of resolving low Mach number flow features accurately.
\end{doublespace}

\begin{doublespace}
\textbf{\large Appendix A: Asymptotic Analysis of  HLLE Scheme for Low Mach Number Flows}
\end{doublespace}

The HLLE scheme shown in (\ref{fv}, \ref{flux-hll}) can be written in non-dimensional form, after rotation,  as 
\begin{equation}
\begin{split}
& |\bar{\Omega}|\dfrac{\partial{}\bar{U}_i}{\partial{}\bar{t}}+  \sum_{\bold{l}\epsilon\upsilon(\bold{i})}\left[\begin{array}{c} \bar{\rho}\bar{u}_n \\ \bar{\rho}\bar{u}\bar{u}_n+\dfrac{\bar{p}}{M_*^2}n_x\\  \bar{\rho}\bar{v}\bar{u}_n+\dfrac{\bar{p}}{M_*^2}n_y \\ (\bar{\rho}\bar{e}+\bar{p})\bar{u_n}\end{array}\right]_{il}+ \left(\dfrac{\bar{u}_nM_*}{2\bar{a}}\right)_{il}\left[\begin{array}{c} \Delta_{il}(\bar{\rho}\bar{u}_n) \\ \Delta_{il}(\bar{\rho}\bar{u}\bar{u}_n)+\left(\dfrac{n_x}{M_*^2}\right)_{il}\Delta_{il}(\bar{p})\\ \Delta_{il}(\bar{\rho}\bar{v}\bar{u}_n)+ \left(\dfrac{n_y}{M_*^2}\right)_{il} \Delta_{il}(\bar{p})\\ \Delta_{il}(\bar{p}\bar{u}_n)\end{array}\right]\Delta{}s_{il}-  \\ 
&  \sum_{\bold{l}\epsilon\upsilon(\bold{i})}\left(\dfrac{\bar{u}_n^2M_*}{2\bar{a}}-\dfrac{\bar{a}}{2M_*}\right)_{il}\left[\begin{array}{c} \Delta_{il}(\bar{\rho}) \\ \Delta_{il}(\bar{\rho}\bar{u})\\ \Delta_{il}(\bar{\rho}\bar{v}) \\ \Delta_{il}(\bar{\rho}\bar{e})\end{array}\right]\Delta{}s_{il}=0
\end{split}
\end{equation}
The flow variables are expanded using the following asymptotic expansions 
\begin{equation} \label{expansion}
\bar{\phi}=\bar{\rho_0}+M_*\bar{\phi_1}+M_*^2\bar{\phi_2}+\mathcal{O}M_*^3
\end{equation}
Expanding the terms asymptotically using equation (\ref{expansion}) and arranging the terms in equal power of $M_*$, we get \\
1. Order of $M_*^0$
\begin{itemize}
\item{Continuity Equation}
\begin{equation}
|\bar{\Omega}|\dfrac{\partial\bar{\rho}_{0i}}{\partial{}\bar{t}}+\sum_{\bold{l}\epsilon\upsilon(\bold{i})}\left((\bar{\rho}_0\bar{u}_{n0})_{il}+\dfrac{\bar{a}_{0il}}{2}\Delta_{il}\bar{\rho}_1\right)\Delta{}s_{il}=0
\end{equation}
\item{x-momentum equation}
\begin{equation}
\begin{split}
& |\bar{\Omega}|\dfrac{\partial(\bar{\rho}_0\bar{u}_0)_i}{\partial{}\bar{t}}+\sum_{\bold{l}\epsilon\upsilon(\bold{i})}\left((\bar{u}_{n0}\bar{\rho}_{0}\bar{u}_0)_{il}+(\bar{p}_2n_x)_{il}+\left(\dfrac{\bar{u}_{n0}}{2\bar{a}_0}n_x\right)_{il}\Delta_{il}\bar{p}_1\right)\Delta{}s_{il} \\ &
+\sum_{\bold{l}\epsilon\upsilon(\bold{i})}\dfrac{\bar{a}_{0il}}{2}\Delta_{il}(\bar{\rho}_0\bar{u}_1+\bar{u}_0\bar{\rho}_1)\Delta{}s_{il}=0
\end{split}
\end{equation}
\item{y-momentum equation}
\begin{equation}
\begin{split}
& |\bar{\Omega}|\dfrac{\partial(\bar{\rho}_0\bar{v}_0)_i}{\partial{}\bar{t}}+\sum_{\bold{l}\epsilon\upsilon(\bold{i})}\left((\bar{u}_{n0}\bar{\rho}_{0}\bar{v}_0)_{il}+(\bar{p}_2n_y)_{il}+\left(\dfrac{\bar{u}_{n0}}{2\bar{a}_0}n_y\right)_{il}\Delta_{il}\bar{p}_1\right)\Delta{}s_{il} \\ &
+\sum_{\bold{l}\epsilon\upsilon(\bold{i})}\dfrac{\bar{a}_{0il}}{2}\Delta_{il}(\bar{\rho}_0\bar{v}_1+\bar{v}_0\bar{\rho}_1)\Delta{}s_{il}=0
\end{split}
\end{equation}
\end{itemize}
2. Order of $M_*^{-1}$ 
\begin{itemize}
\item{Continuity Equation}
\begin{equation}
\sum_{\bold{l}\epsilon\upsilon(\bold{i})}\dfrac{1}{2}a_{0,il}\Delta_{il}\bar{\rho}_0\Delta{}s_{il}=0
\end{equation}
Here $\sum_{\bold{l}\epsilon\upsilon(\bold{i})}\Delta_{il}\bar{\rho}_0\Delta{}s_{il}=0$ and hence $\bar{\rho}_0=$Constant.

\item{x-momentum equation}
\begin{equation} \label{x-mom}
 \sum_{\bold{l}\epsilon\upsilon(\bold{i})}\left((\bar{p}_1n_x)_{il}+\left(\dfrac{\bar{u}_{n0}}{2\bar{a}_0}n_x\right)_{il}\Delta_{il}\bar{p}_0+\dfrac{\bar{u}_{0il}}{2\bar{a}_{0il}}\Delta_{il}(\bar{p}_0)+\dfrac{a_{0il}}{2}\Delta_{il}(\bar{\rho}_0\bar{u}_0)\right)\Delta{}s_{il}=0
\end{equation}

\item{y-momentum equation} 
\begin{equation} \label{y-mom}
 \sum_{\bold{l}\epsilon\upsilon(\bold{i})}\left((\bar{p}_1n_y)_{il}+\left(\dfrac{\bar{u}_{n0}}{2\bar{a}_0}n_y\right)_{il}\Delta_{il}\bar{p}_0+\dfrac{\bar{v}_{0il}}{2\bar{a}_{0il}}\Delta_{il}(\bar{p}_0)+\dfrac{a_{0il}}{2}\Delta_{il}(\bar{\rho}_0\bar{v}_0)\right)\Delta{}s_{il}=0
\end{equation}

\item{energy equation}
\begin{equation}
\sum_{\bold{l}\epsilon\upsilon(\bold{i})}\dfrac{1}{2}a_{0,il}\Delta_{il}(\bar{\rho}_0\bar{e}_0)\Delta{}s_{il}=0
\end{equation}
\end{itemize}

From the above equation we obtain $\sum_{\bold{l}\epsilon\upsilon(\bold{i})}\Delta_{il}\bar{p}_0\Delta{}s_{il}=0$. This implies that $\bar{p}_0=cte\forall{}\bold{i}$ where $cte$ is a constant.

Therefore, equations (\ref{x-mom},\ref{y-mom}) reduces to 
\begin{equation}
 \sum_{\bold{l}\epsilon\upsilon(\bold{i})}(\bar{p}_1n_x)_{il}=-\dfrac{a_{0il}}{2}\Delta_{il}(\bar{\rho}_0\bar{u}_0) \hspace {2cm}  \sum_{\bold{l}\epsilon\upsilon(\bold{i})}(\bar{p}_1n_y)_{il}=-\dfrac{a_{0il}}{2}\Delta_{il}(\bar{\rho}_0\bar{v}_0)
\end{equation}

This implies that $\bar{p}_1\neq{}cte\forall{}\bold{i}$ where $cte$ is a constant.  Therefore, the  HLLE scheme for the discrete Euler equations supports pressure fluctuation of the type $p(x,t)=P_0(t)+M_*p_1(x,t)$ and hence shall not be able to resolve the  low Mach flow features. It can be seen from the above equation that the velocity jump in the normal direction is a cause of the unphysical behaviour of the HLLE scheme in low Mach flows.

\end{document}